\documentclass[a4paper,fleqn,usenatbib,useAMS]{mnras}

\usepackage[T1]{fontenc}
\usepackage{ae,aecompl}

\usepackage{graphicx}	
\usepackage{amsmath}	
\usepackage{amssymb}	
\usepackage{multirow}
\usepackage{longtable}

\usepackage{graphicx}	
\usepackage{multicol}        
\usepackage{bm}		
\usepackage{pdflscape}	
\usepackage{threeparttable}

\title[Shocked and ionized gas in M17]{Spatial distributions and kinematics of shocked and ionized gas in M17}

\author[F.-Y. Zhu et al.]{
Feng-Yao Zhu,$^{1,2}$\thanks{E-mail: zhufy@zhejianglab.com}
Junzhi Wang,$^{3}$\thanks{E-mail: junzhiwang@gxu.edu.cn}
Yaoting Yan,$^{4}$
Qing-Feng Zhu$^{5}$
Juan Li$^{2}$
\\
$^{1}$Research Center for Intelligent Computing Platforms, Zhejiang Laboratory, Hangzhou, 311100, PR China\\
$^{2}$Shanghai Astronomical Observatory, Chinese Academy of Sciences, Shanghai, 200030, PR China\\
$^{3}$Guangxi Key Laboratory for Relativistic Astrophysics, School of Physical Science and Technology, \\
Guangxi University, Nanning 530004, PR China\\
$^{4}$Max-Planck-Institut f\"{u}r Radioastronomie, Auf dem H\"{u}gel 69, 53121 Bonn, Germany\\
$^{5}$Astronomy department, University of Science and Technology of China, Hefei, 230026, PR China\\
}

\date{Accepted XXX. Received YYY; in original form ZZZ}

\pubyear{2020}

\begin{document}
\label{firstpage}
\pagerange{\pageref{firstpage}--\pageref{lastpage}}
\maketitle

\begin{abstract}
Massive stars are formed in molecular clouds, and produce H II regions when they evolve onto the main sequence. The expansion of H II region can both suppress and promote star formation in the vicinity. M17 H II region is a giant cometary H II region near many massive clumps containing starless and protostellar sources. It is an appropriate target to study the effect of feedback from previously formed massive stars on the nearby star-forming environments. Observations of SiO 2-1, HCO$^+$ 1-0, H$^{13}$CO$^+$ 1-0, HC$_3$N 10-9, and H41$\alpha$ lines are performed toward M17 H II region with ambient candidates of massive clumps. In the observations, the widespread shocked gas surrounding M17 H II region is detected: it probably originates from the collision between the expanding ionized gas and the ambient neutral medium. Some massive clumps are found in the overlap region of the shock and dense-gas tracing lines while the central velocities of shocked and high-density gases are similar. This suggests that part of massive clumps are located in the shell of H II region, and may be formed from the accumulated neutral materials in the shell. In addition, by comparing the observations toward M17 H II region with the simulation of cometary H II region, we infer the presence of one or more massive stars travelling at supersonic velocity with respect to the natal molecular cloud in the H II region.
\end{abstract}




\section{Introduction} \label{sec:intro}

Massive stars play important role in the evolution of galaxies \citep{mot18}. They first form in molecular clouds, and then affect the parental clouds by the feedbacks including stellar wind, ionizing and dissociating radiation. When massive stars evolve into main-sequence stage, the large number of extreme-ultraviolet photons ($h\nu\ge13.6$ eV) from the massive star can ionize the ambient interstellar medium (ISM). The region composed of ionized gas is called H II region. The ionized gas is significantly heated by the ionizing radiation so that the temperature and pressure of the ionized gas become much higher than those of the nearby neutral gas. This leads to the expansion of H II region, and the swept-up surrounding neutral materials form a shell around the ionized gas. With more and more molecular gas accumulated, the gravitational fragments are expected to form in this shell \citep{hos06}. In this way, massive stars can promote the star formation in the vicinity. A number of previous observations showed evidences for star formation caused by expanding H II regions \citep{jia02,kar03,ker13,pov09,shi19}. However, the dense molecular cloud which could lead to star formation may also be destroyed by ionizing radiation. When H II regions expand beyond the boundary of clouds, a large mass of dense gas will leave from clouds through the fast flowing of ionized gas. This can suppress the star formation in the parental clouds of H II regions \citep{hos06}.

M17 is a star-forming region in the galaxy with extensive studies \citep{sof22}. There is a giant cometary H II region with a young age of $\le1.0$ Myr, a $\sim 5'$ diameter, and a luminosity of $3.6\times10^6$ L$_\odot$ \citep{han97,kuh19,bor22} in M17. At least 16 O stars and 100 B stars are located in the H II region, and they compose the massive cluster, NGC 6618 \citep{han97,hof08,pov09}. The O stars contribute the majority of ionizing radiation and stellar wind materials. M17 H II region is half surrounded in a molecular cloud \citep{wil99,pov09}. A dense molecular cloud core M17 SW is located close to the southwestern boundary of the H II region. It is highly clumpy and dense enough for gravitational collapse \citep{lad76,stu90}. Many protostellar candidates have also been found in the molecular cloud near M17 H II region. A part of them are thought to be triggered by the massive stars in M17 H II region \citep{hof08}.

A large sample of massive clumps were detected in the 1.1 mm continuum Bolocam Galactic Plane Survey (BGPS) \citep{gin13}. In this sample, 2223 starless clump candidates (SCCs) and 2460 protostellar sources were identified \citep{svo16}. \citet{zhu20} performed observations of the SiO 1-0, 2-1 and 3-2 lines toward 100 SCCs selected from the starless sample given in \citet{svo16}. Among the 100 SCCs in \citet{zhu20}, 4 sources are located near M17 H II region. Shocks indicated by the SiO lines were detected in the 4 sources. Although these shocks could be originated from the outflows of intermediate and low-mass protostars deeply embedded in these SCCs, the expansion of H II region can also produce shocks \citep{art06,zhu15b}. The origins of the shocks are unknown. If these shocks result from the expansion of H II region, the formation of these starless clumps may also be related with M17 H II region. Additionally, there are a number of starless and protostellar clumps near M17 H II region. Studying the interaction between M17 H II region and these clumps is conducive to learning the effects of H II regions on the environment of nearby star formation.   

Observations of hydrogen recombination lines (RRLs) and molecular lines toward M17 H II region and the nearby molecular cloud containing massive clumps can help us to study the properties of the ionized and neutral gas including temperature, density, and kinematics. These properties can also reveal the information about the interaction between the H II region and surrounding molecular clumps. The relations between M17 H II region and the nearby starless and protostellar clumps could then be studied. \citet{elm77} suggested that the ionization-front shocks driven into molecular clouds by previously formed massive stars can trigger new star formation. \citet{hof08} also proposed that star formation could be enhanced at the interface of M17 H II region with the molecular cloud. This effect can be checked by the relations between M17 H II region and nearby clumps.


In this work, we present the mapping observations of the SiO 2-1, HCO$^+$ 1-0, H$^{13}$CO$^+$ 1-0, HC$_3$N 10-9 and H41$\alpha$ lines toward M17 H II region and the ambient interstellar medium (ISM). The distribution of shocked gas is shown by the SiO 2-1 line \citep{gus08}. The HCO$^+$, H$^{13}$CO$^+$ 1-0, HC$_3$N 10-9 can trace high-density gas \citep{liu20}. The H41$\alpha$ line is used to trace ionized gas \citep{woo89}. The property distributions of the molecular and ionized gas are shown. The velocity field of the ionized gas is studied. The formation of the cometary H II region and the effect of the H II region on the surrounding massive clumps are both discussed. The organization of this work is as follows. The information about the observations toward M17 H II region with ambient ISM is provided in Section \ref{sec:method}. In Section \ref{sec:result}, the results of the observations and the relevant analyses are presented. The discussions about the formation of the cometary morphology of M17 H II region and the relation between M17 H II region and nearby clumps are written in Section \ref{sec:discussion}. The summary and conclusion are shown in Section \ref{sec:conclusion}.

\section{Observations} \label{sec:method}

\subsection{M17 H II region and surrounding massive clumps}

The 1.06 GHz continuum map of M17 H II region given in \citet{beu16} is presented in Figure \ref{fig:obs_fields}. As mentioned above, M17 H II region is a giant and cometary H II region. The head of M17 H II region is at northwest, and the broad tail is at southeast. There is a bubble composed of stellar wind material with an extremely high temperature $T_e\approx7\times10^6$ K and low density $n_e\approx0.3$ cm$^{-3}$ in the center of M17 H II region \citep{tow03,pov08}. The photoionized region with relatively lower temperature $T_e\sim6600-9100$ K and higher electron density $n_e\approx10^3$ cm$^{-3}$ is around the bubble \citep{hje71,dow80,sub96,joh98,tow03}. The distance to M17 seems to be under debate \citep{hof08}. Different values of distance to M17 ranging from 1.3 to 2.9 kpc have been presented using different measuring methods \citep{han97,nie01}. A distance of 1.98$^{+0.14}_{-0.12}$ kpc is estimated from the motion of methanol masers in M 17 \citep{xu11}.

There are 11 SCCs and 5 protostellar clumps near M17 H II region covered in the current observations. These sources are found in the 1.1 mm continuum BGPS, identified as dense gas clumps in \citet{gin13}, and classified as massive starless and protostellar candidates in \citet{svo16}. Four of the 11 SCCs, BGPS 3110, 3114, 3118, and 3128, were observed in our previous single-dish observations and all detected in SiO lines \citep{zhu20}. 





\subsection{Observations and Data reduction}

\begin{figure}
  \centering
  \includegraphics[scale=0.5]{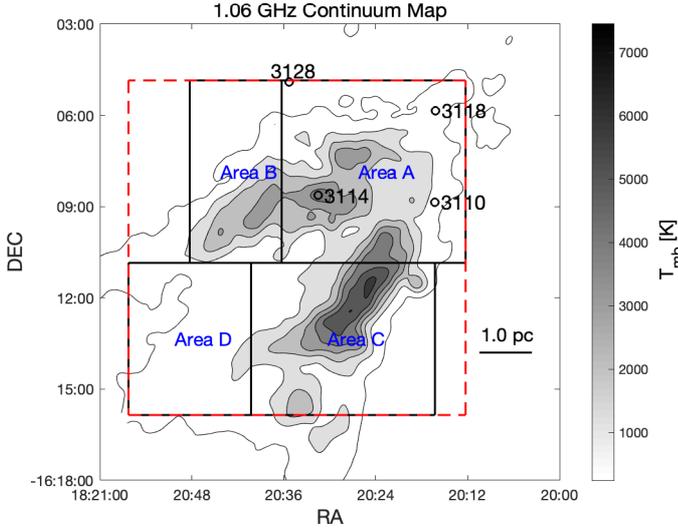}
  \caption{ The 1.06 GHz continuum map around M17 H II region from the THOR project using VLA \citep{beu16}. The positions of the SCCs detected in the SiO lines are shown as black circles. The black solid lines indicates the 4 small areas composing the whole observational field. The red dashed lines show the extent of the image presented in Figure \ref{fig:H41a_ctn}.}\label{fig:obs_fields}
\end{figure}

We performed mapping observations of hydrogen recombination lines and molecular lines toward M17 H II region using IRAM 30-m telescope from December 2020 to January 2021. We used 3 mm (E0) band of the Eight Mixer Receivers (EMIR) receiver and the FTS backend with a 195 kHz channel width and an 8 GHz band width for the observations. The typical spectrometer resolution is 0.67 km s$^{-1}$ in the 3 mm band. The OTF PSW observing mode was used to observe most part of the H II region. The observational field covered in our observations are shown in Figure \ref{fig:obs_fields}. The whole field is divided into 4 small parts in our observations. The beam size is $\sim28''$ for the 3 mm band. The reference positions are 1 degree offset from the centers of each small observation fields in RA. The system temperature is typically 100 K. The line frequencies, the beam sizes and the main-beam efficiencies corresponding to the observed molecular and atomic lines are listed in Table \ref{table_line}.


\begin{table} \tiny 
\centering
\caption{The frequencies, beam sizes, main beam efficiencies, and fluxes per Kelvin on the $T^*_A$ scale corresponding to the observed molecular transitions for IRAM 30m telescope.}\label{table_line}
\begin{tabular}{|ccccc|}
\hline
Transition & Frequency [MHz] & Beam size & B$_{eff}$ & S$_\nu$/$T^*_A$ [Jy/K] \\
\hline
SiO 2-1 & 86847.00 & 28$''$ & 0.81 & 5.8 \\
HCO$^+$ 1-0 & 89188.53 & 28$''$ & 0.81 & 5.8 \\
H$^{13}$CO$^+$ 1-0 & 86754.29 & 28$''$ & 0.81 & 5.8 \\
HC$_3$N 10-9 & 90979.02 & 27$''$ & 0.81 & 5.8 \\
H41$\alpha$ & 92034.43 & 27$''$ & 0.81 & 5.8 \\
\hline
\end{tabular}
\end{table}

The observational data were reduced with GILDAS software \citep{gui88} to obtain the spectra of the observed lines in individual pixels. Linear baselines were removed from the spectra. The rms noise level in T$_{mb}$ is typically 33 mK per pixel in the 3 mm band. The noise level in Area C is slightly increased by about 15$\%$ because of less observation time. Near the boundaries of each small observational fields, the noise level increases by about $40\%$.

\section{results} \label{sec:result}

In this section, the results of observations of the H41$\alpha$, SiO 2-1, HCO$^+$ 1-0, H$^{13}$CO$^+$ 1-0, and HC$_3$N 10-9 lines toward M17 H II region with the ambient medium are presented. The distributions of the ionized and molecular gas are then studied. The 1.06 GHz continuum observed by THOR given in \citet{beu16} is also used in the relevant analyses.

In the Table \ref{table:global_property}, the global properties of these lines from the whole observational field are provided. They are estimated from line emissions integrated over the whole observational field by using single Gaussian fitting. Although the gas components indicated by these atomic and molecular lines are different, the central velocities of these lines are all approximately close to each others. This is consistent with the conclusion in previous works that the rest velocity of the molecular cloud is $\sim20$ km s$^{-1}$ \citep{pov09,sof22}.

The central velocity of H41$\alpha$ line emitted from ionized gas is not significantly different from the velocity of neutral gas indicated by the molecular lines. This is not very surprising although M17 H II region has a cometary morphology. From previous simulations of cometary H II regions \citep{zhu15b}, the central velocity of the line emission spatially integrated from the whole H II region could still be less than 5 km s$^{-1}$ different from the rest velocity even when there is a strong density gradient or a high stellar velocity.

On the contrary, the width of H41$\alpha$ line is much different from those of molecular lines. This should be because of the higher temperature, and stronger velocity field of ionized gas than those of molecular gas in the large scale. 

\begin{table} \tiny 
\centering
\caption{The global properties of the observed lines including the velocity-integrated flux density ($\int S_\nu dv$), central velocity, and FWHM.}\label{table:global_property}
\begin{tabular}{|c|ccc|}
\hline
Transition & $\int S_\nu dv$ [$10^3$ Jy km s$^{-1}$] & $v$ [km s$^{-1}$] & FWHM [km s$^{-1}$] \\
\hline
SiO 2-1 & $0.63\pm0.02$ & $20.0\pm0.1$ & $7.6\pm0.3$ \\
HCO$^+$ 1-0 & $22.8\pm0.2$ & $18.7\pm0.0$ & $7.1\pm0.1$ \\
H$^{13}$CO$^+$ 1-0 & $1.21\pm0.02$ & $19.5\pm0.1$ & $4.9\pm0.1$ \\
HC$_3$N 10-9 & $2.16\pm0.03$ & $18.4\pm0.0$ & $4.5\pm0.1$ \\
H41$\alpha$ & $9.73\pm0.04$ & $18.9\pm0.1$ & $33.6\pm0.2$ \\
\hline
\end{tabular}
\end{table}

\subsection{Ionized gas shown in the H41$\alpha$ line}

\begin{figure}
  \centering
  \includegraphics[scale=0.5]{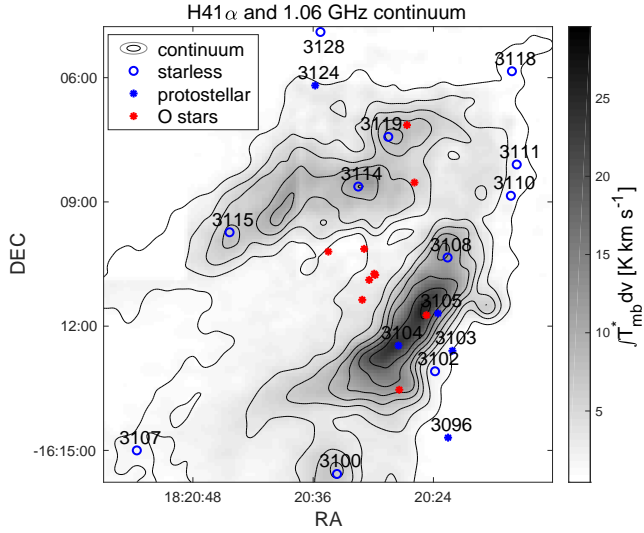}
  \caption{ The comparison of the distributions of the H41$\alpha$ line and 1.06 GHz continuum intensity toward M17. The contour levels start at 100 K in steps of 800 K of the continuum $T_{mb}$. The velocity-integrated intensity of H41$\alpha$ line is shown by the gray-scale image. The blue circles and asterisks are the positions of starless and protostellar clump candidates identified in \citet{svo16}, respectively. The red asterisks are the positions of the massive stars earlier than O9 V \citep{hof08}.}\label{fig:H41a_ctn}
\end{figure}

The distributions of velocity-integrated intensity ($\int T_{mb}dv$) of the H41$\alpha$ line and intensity ($T_{mb}$) of 1.06 GHz continuum are presented in Figure \ref{fig:H41a_ctn}. The distribution of H41$\alpha$ line is similar to that of 1.06 GHz continuum. There is an arch-like region with compact emission both of line and continuum in M17 H II region. The two sides of the compact emission are bar-like features which were called southern and northern bars in previous studies, respectively \citep{bro01,jia02}. The typical values of the velocity-integrated H41$\alpha$ line intensity and the continuum intensity are 7.5 K km s$^{-1}$ and 3000 K in the northern bar, respectively. And those in the southern bar are 12 K km s$^{-1}$ and 4000 K. The southern bar of M17 H II region is brighter than the northern bar both in 1.06 GHz continuum and H41$\alpha$ line. It is easy to understand because the southern bar is closer to M17 SW, the core of molecular cloud. The expansion of ionized gas is limited in this direction so that the ionized gas is more compact in the southern bar than in the northern bar. Moreover, other parts of the H II region are relatively fainter. The velocity-integrated H41$\alpha$ line intensity is about 3.0 K km s$^{-1}$ in the apex surrounded by BGPS 3108, 3110, 3118, and 3119, and is about 1.0 K km s$^{-1}$ in the tail near BGPS 3107. In addition, the distribution of the X-ray-emitting OB stellar winds shown by \citet{tow03} roughly matches the faint region between the northern and southern bars presented in the H41$\alpha$ line. The stellar wind has little contribution to hydrogen RRLs because of very high temperature and low density.

\begin{figure}
  \centering
  \includegraphics[scale=0.5]{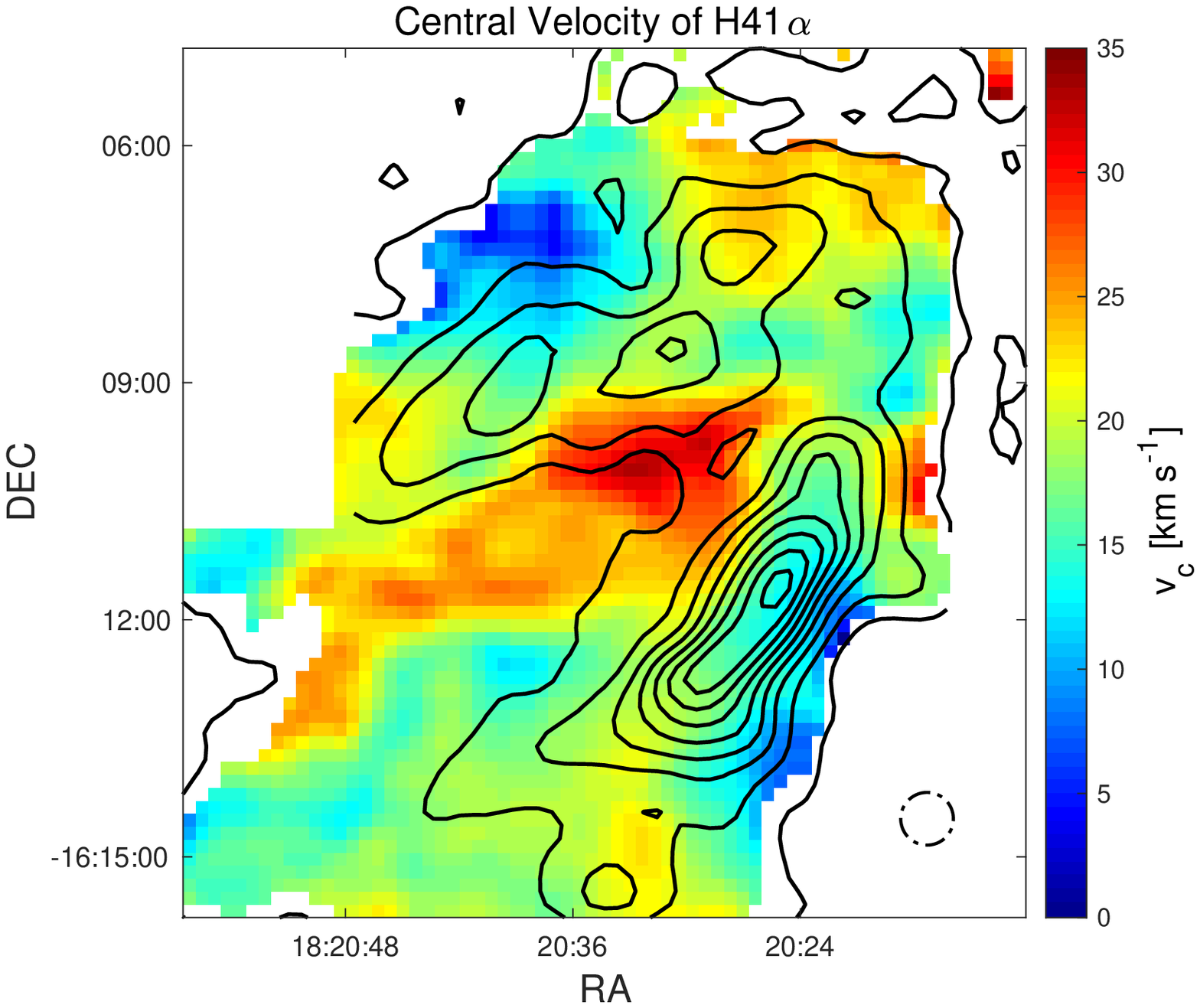}
  \includegraphics[scale=0.5]{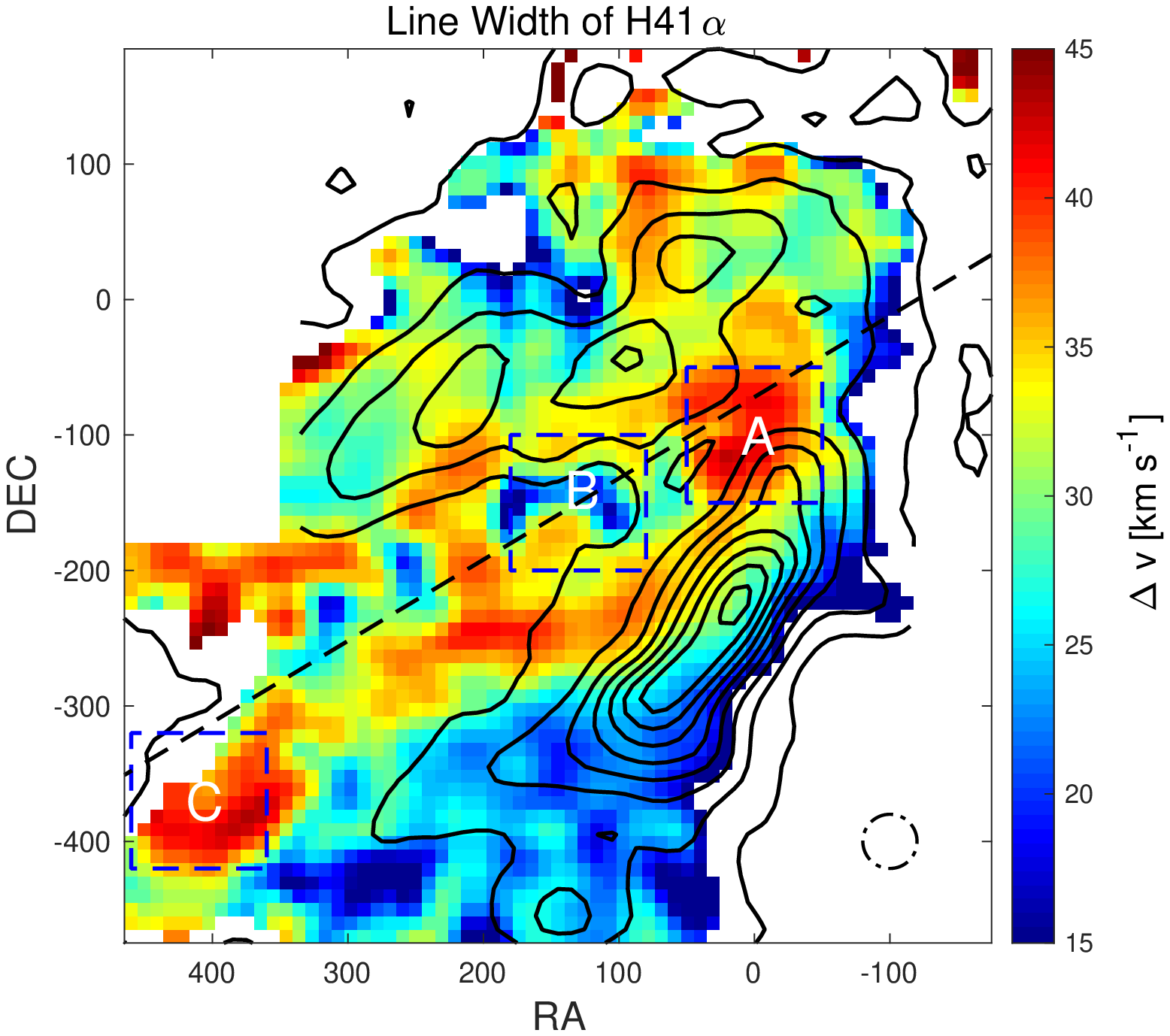}
  \caption{Distributions of central velocity and width of the H41$\alpha$ line shown in the top and bottom panels, respectively. The contour levels starts at 3 $\sigma$ in steps of 12 $\sigma$ of $\int T_{mb}dv$ of H41$\alpha$ line. 1 $\sigma$ is 0.25 K km s$^{-1}$. The dash-dotted circles show the spatial resolution of $40''$. The black dashed line in the bottom panel is the slit from which the P-V diagram is presented in Figure \ref{fig:H41a_pvdiagram}. The velocity range of integration is from -20 km s$^{-1}$ to 60 km s$^{-1}$.}\label{fig:H41a_center_width}
\end{figure}

The distributions of the flux weighted central velocity ($v_c$) and width of the H41$\alpha$ line are displayed in Figure \ref{fig:H41a_center_width}. In order to exclude the false values due to noise, the information of H41$\alpha$ line is neglected in the area where the signal-to-noise ratio is lower than 5. And the spatial resolution is reduced to be about $40''$. In the top panel, A zone composed of red-shifted ionized gas is distributed from the center to the southeast of the H II region. The position of the O stars between southern and northern bars is defined as the center of H II region in this work. The gas velocity is highest in the center region with $v_c=34.0$ km s$^{-1}$, and it gradually decreases to be about 20 km s$^{-1}$ in the tail. The distribution area of the red-shifted gas seems to correspond to the distribution of stellar wind shown in Figure 3b of \citet{tow03}.

The line width presented in the bottom panel of Figure \ref{fig:H41a_center_width} is defined as below,

\begin{equation}
\Delta v=2[2ln2\frac{\int T_{mb}(v-v_c)^2dv}{\int T_{mb}dv}]^{1/2}~~~.
\end{equation}

\begin{figure}
  \centering
  \includegraphics[scale=0.4]{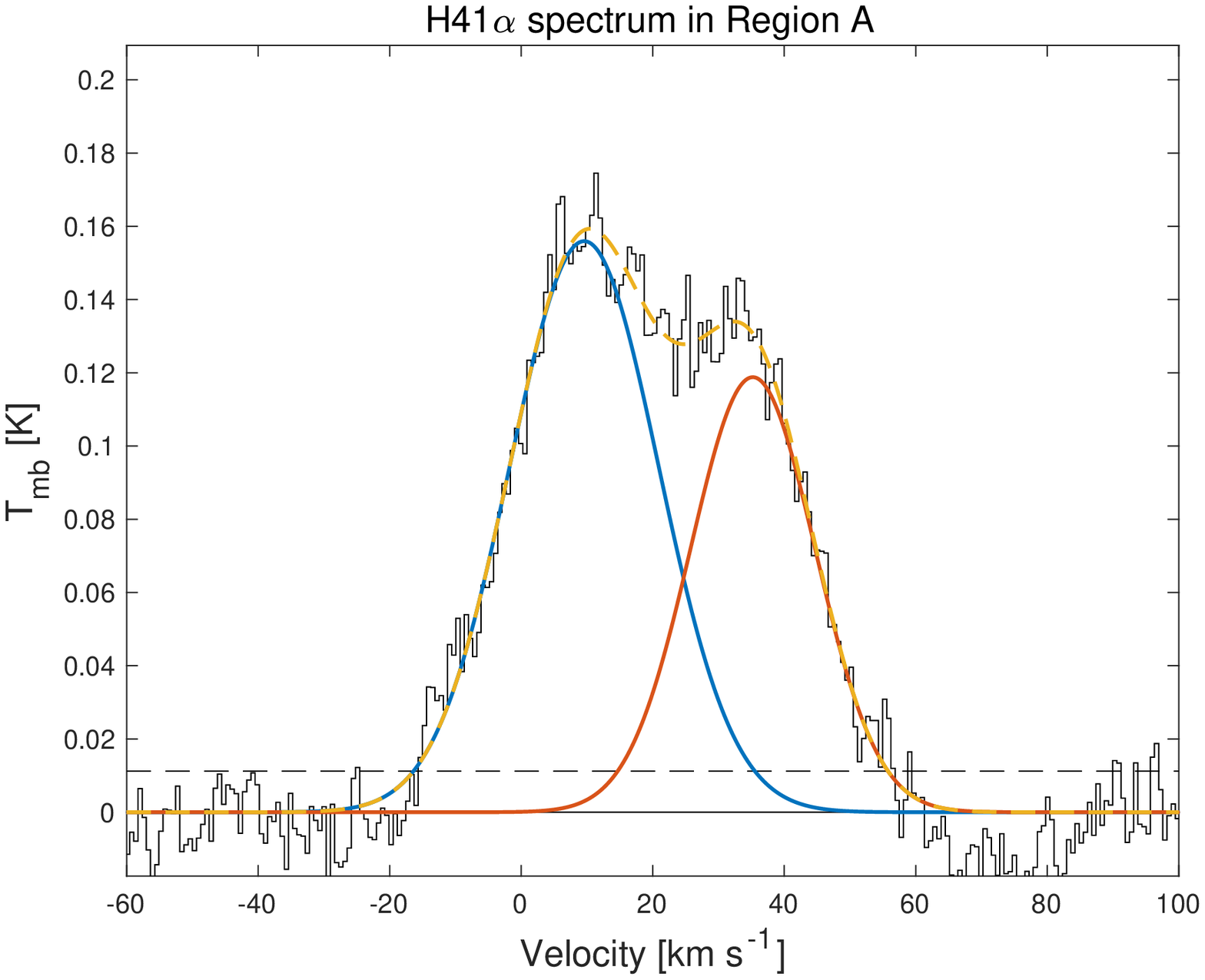}
  \includegraphics[scale=0.4]{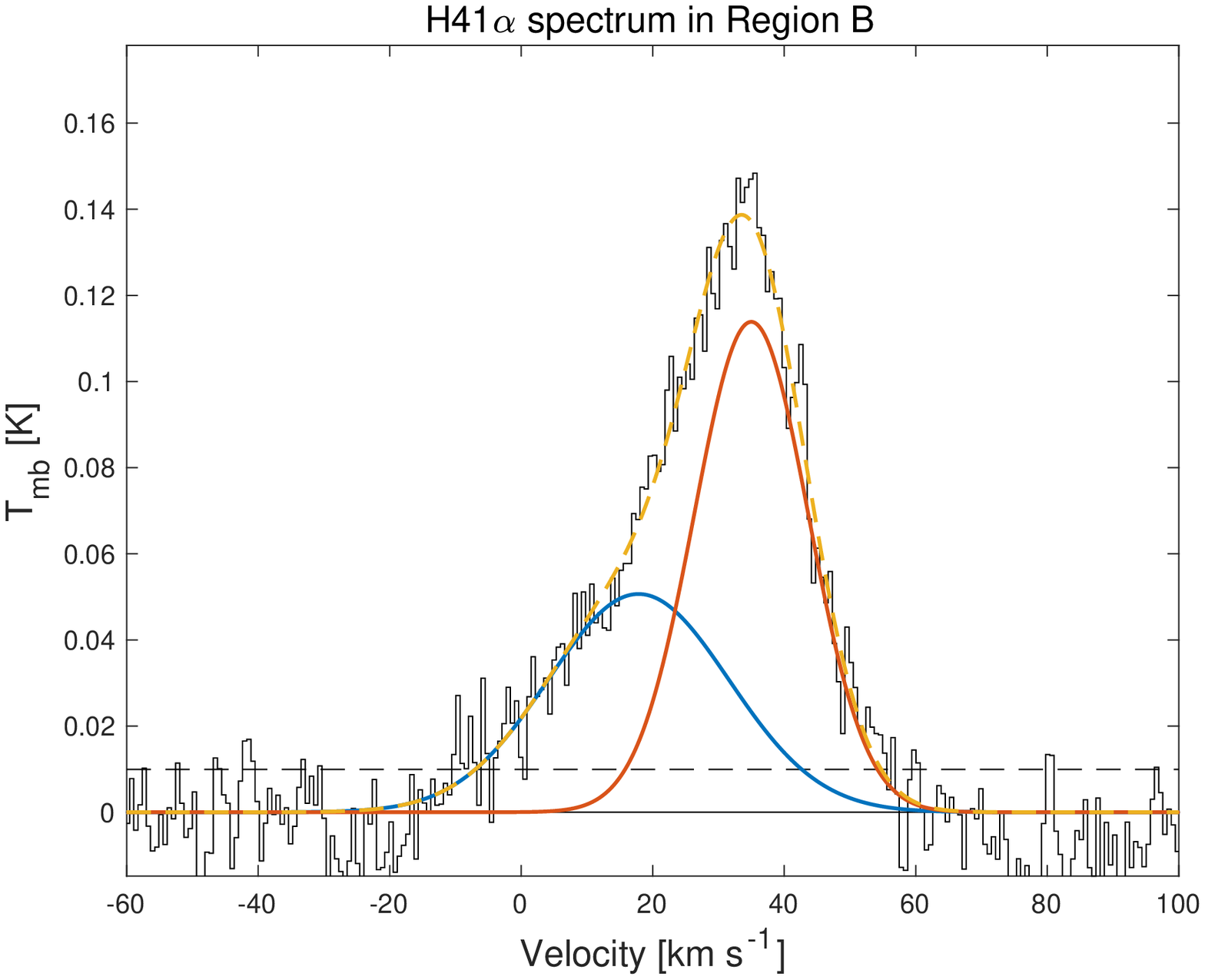}
  \includegraphics[scale=0.4]{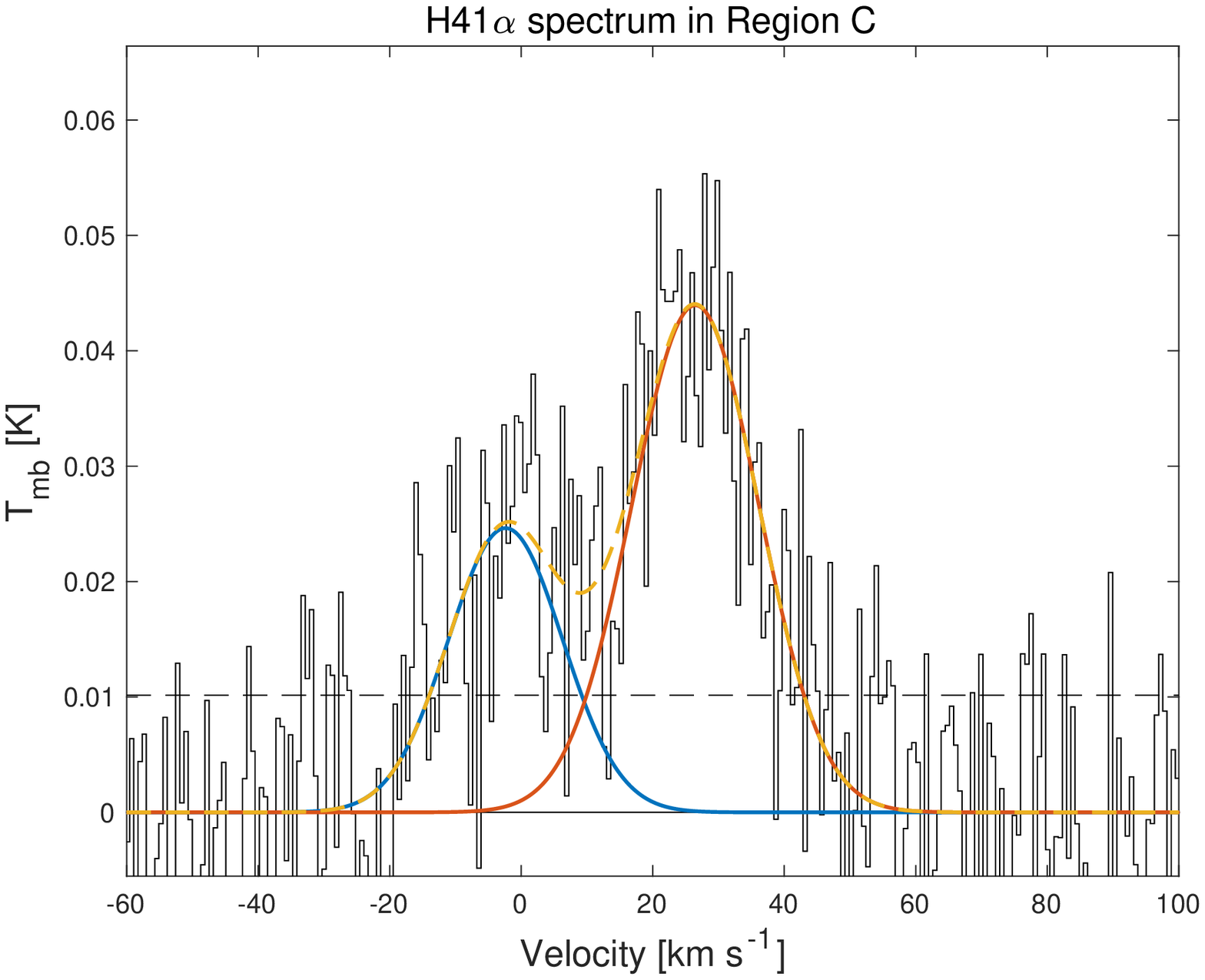}
  \caption{H41$\alpha$ spectra toward Regions A, B, and C marked in the bottom panel of Figure \ref{fig:H41a_center_width}. The black stairs mean the H41$\alpha$ spectrum. The blue and red curves are the gaussian curves fitting the contributions from two velocity components of ionized gas.}\label{fig:H41a_spectrum}
\end{figure}

The line width $\Delta v$ is equal to FWHM when the line profile is Gaussian. It is interesting that a layer of ionized gas with broad H41$\alpha$ line width envelops the red-shifted gas. The spectra of H41$\alpha$ line toward 2 positions (Regions A and C) in the layer and 1 position (Region B) in the red-shifted zone are plotted in Figure \ref{fig:H41a_spectrum}. All of these spectra can be divided into two gaussian components with a red-shifted velocity and a blue-shifted velocity. For the spectra toward Regions A and C, the contributions of the two components are both significant. On the contrary, the blue-shifted component is not important in the H41$\alpha$ spectrum toward Region B.

Compared with the simulations of cometary H II regions with a stellar wind bubble \citep{zhu15b}, these two velocity components in the H41$\alpha$ spectrum should correspond to the ionized gas in the near and father sides of photoionized region divided by the stellar wind bubble along the line of sight. Since M17 H II region is a cometary H II region observed from the tail to the head with an inclination angle, the ionized gas in the near side along the line of sight is closer to the tail of H II region than that in the father side. And in a cometary H II region, photoionized gas closer to the tail shows more blue-shifted velocity. This leads to the double-Gaussian profiles in the H41$\alpha$ spectra.

\begin{figure}
  \centering
  \includegraphics[scale=0.5]{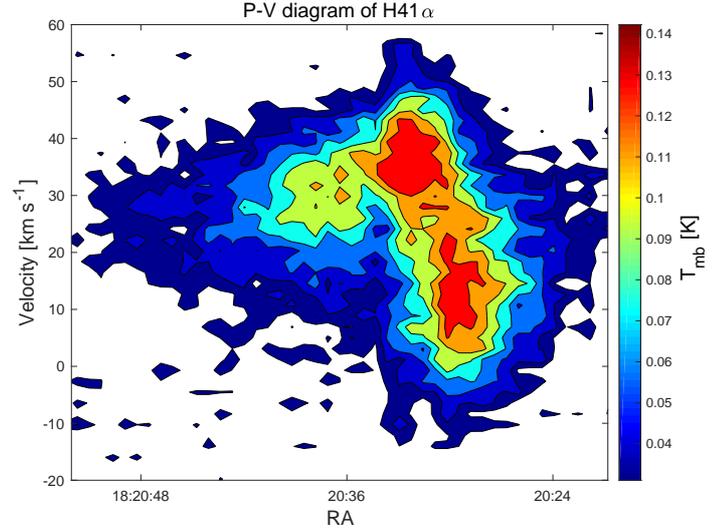}
  \caption{Position-velocity diagram of the H41$\alpha$ line from the slit along the dashed line in the bottom panel of Figure \ref{fig:H41a_center_width}. }\label{fig:H41a_pvdiagram}
\end{figure}

The position-velocity diagram (P-V diagram) from the slit roughly along the axis of the stellar wind bubble in M17 H II region is presented in Figure \ref{fig:H41a_pvdiagram}. It is similar to those for the [Ne II] 12.81 $\mu$m line in a bow-shock model presented in Figure 5 of \citet{zhu15b}. Although the width of [Ne II] 12.81$\mu$m line is commonly narrower than that of H41$\alpha$ line, the similar velocity distributions of ionized gas can still be recognized in the P-V diagrams.

\subsection{The HCO$^+$, H$^{13}$CO$^+$ 1-0, HC$_3$N lines tracing dense gas}

\begin{figure}
  \centering
  \includegraphics[scale=0.45]{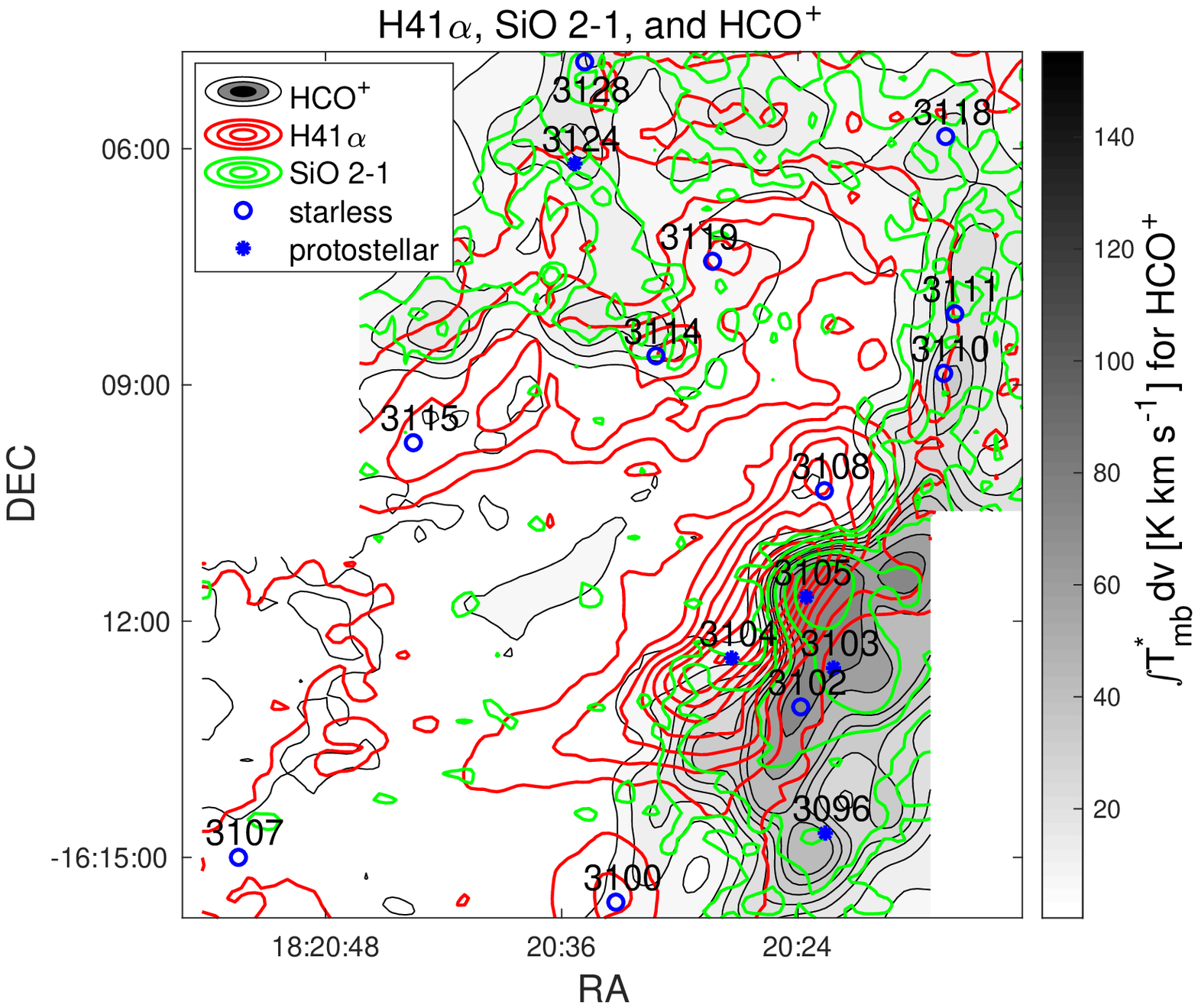}
  \includegraphics[scale=0.45]{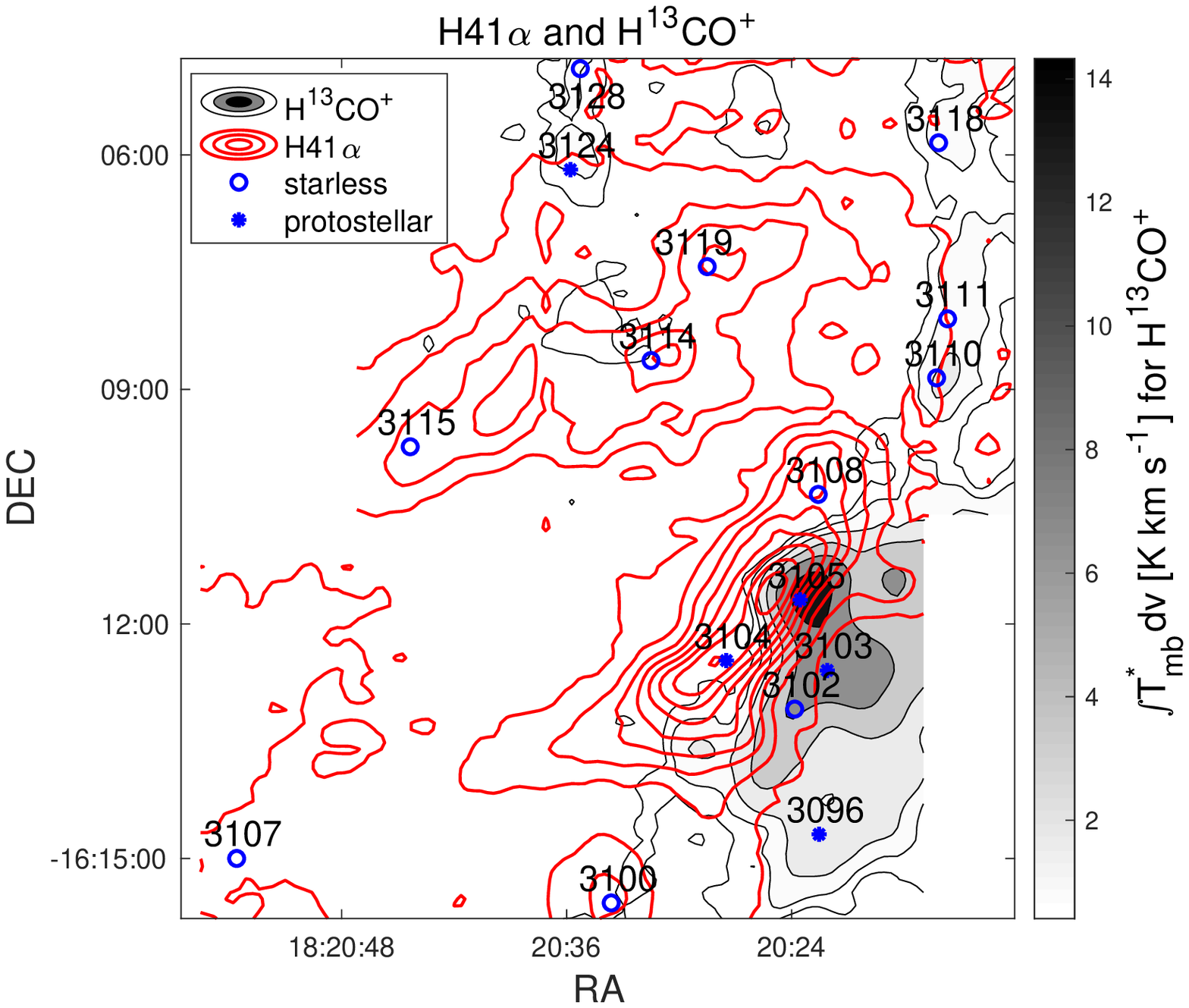}
  \includegraphics[scale=0.45]{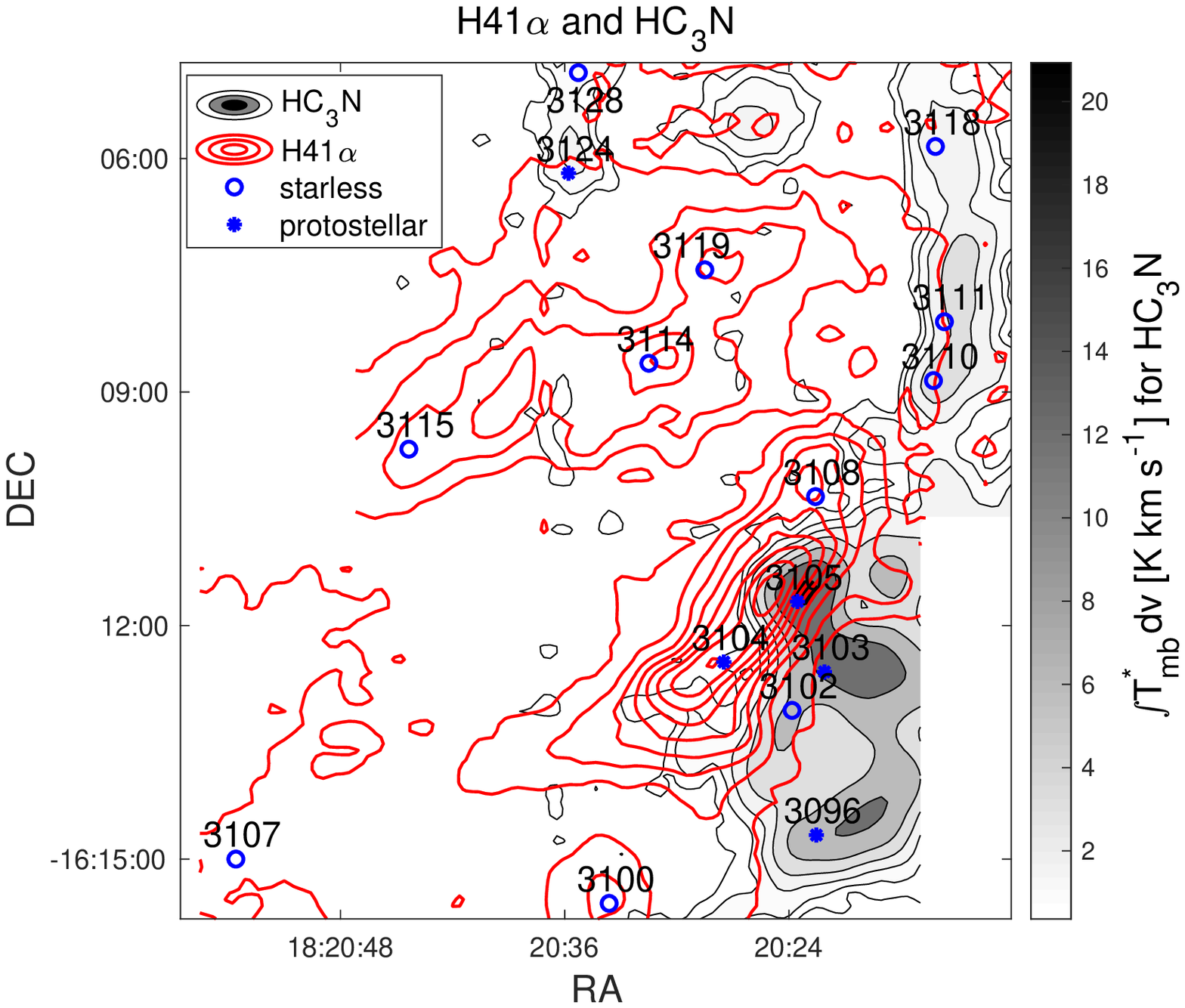}
  \caption{Distributions of velocity-integrated intensity of HCO$^+$ 1-0, H$^{13}$CO$^+$ 1-0, HC$_3$N 10-9, SiO 2-1 and H41$\alpha$ lines. The positions of starless and protostellar clump candidates are indicated by blue circles and asterisks, respectively. The red contour levels of H41$\alpha$ emission start at 5$\sigma$ in steps of 10$\sigma$ with 1$\sigma=0.31$ K km s$^{-1}$. The green contour levels of SiO 2-1 emission starts at 2$\sigma$, and the nth level indicates the value of 2$^n\sigma$ with 1$\sigma=0.13$ K km s$^{-1}$.}\label{fig:HCO+_H13CO+}
\end{figure}

The intensity distributions of HCO$^+$ and H$^{13}$CO$^+$ 1-0 lines are presented in Figure \ref{fig:HCO+_H13CO+}. The distribution of H41$\alpha$ line is also presented to study the spatial correlations between the dense and ionized gas. The HCO$^+$ 1-0 line emission is widespread in the whole observational field. However, the line emission is much brighter near the boundary of H II region than in the region overlapped by the H II region in the projected image. The H$^{13}$CO$^+$ 1-0 line is weaker but more sensitive to the high-density gas than the HCO$^+$ 1-0 line because the HCO$^+$ line is normally optically thick in star-forming regions \citep{cal18,liu20}. But the distributions of these two lines both indicate that a dense layer of molecular gas spreads along the boundary of M17 H II region at the west and north directions. The intensities of the two lines are high in the layer and decrease on both sides. The brightest emission in both the HCO$^+$ and H$^{13}$CO$^+$ 1-0 lines is from the area near BGPS 3105 corresponding to the northern condensation in cloud core M17 SW \citep{wan93}.


The central velocities of HCO$^+$ and H$^{13}$CO$^+$ 1-0 lines have not significant variation with the position in the observation field. They are respectively 19.2 km s$^{-1}$ and 19.7 km s$^{-1}$ toward the high-density region near BGPS 3103 and 3105, and about 19.0 km s$^{-1}$ and 18.5 km s$^{-1}$ toward BGPS 3124. The H$^{13}$CO$^+$ 1-0 line is not detected toward BGPS 3107, but the HCO$^+$ 1-0 line is detected with a central velocity of 19.4 km s$^{-1}$.

In addition, the distribution of HC$_3$N 10-9 emission is presented in the bottom panel of Figure \ref{fig:HCO+_H13CO+}. Although the excitation temperature and critical density of HC$_3$N 10-9 line is higher than those of HCO$^+$ and H$^{13}$CO$^+$ 1-0 lines \citep{sch05,shi15}, the distribution of  HC$_3$N 10-9 line is roughly similar to the distribution of H$^{13}$CO 1-0 line. The HC$_3$N emission is mainly distributed along the boundary of M17 H II region at the west and north directions. It is also brightest in the northern condensation of M17 SW as the HCO$^+$ and H$^{13}$CO$^+$ lines are. The similarity of the distributions of the HC$_3$N and H$^{13}$CO$^+$ lines suggests that these two lines are reliable tools to trace dense gas since the intensity distributions of the lines are mainly determined by the high-density gas distribution rather than the distribution of chemical abundance.


\subsection{The shocked gas indicated by the SiO 2-1 line} \label{sec:SiO2-1}

\begin{figure}
  \centering
  \includegraphics[scale=0.5]{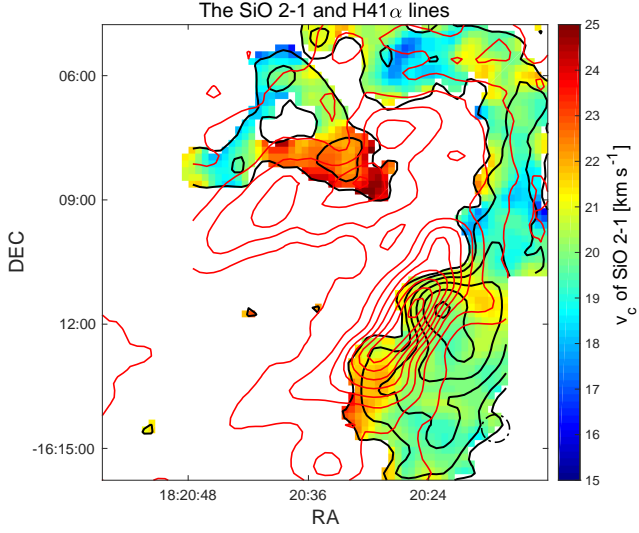}
  \caption{Distributions of velocity-integrated intensity and central velocity of SiO 2-1 and H41$\alpha$ lines. The black contour levels of the SiO 2-1 line indicate the line intensity value of 2$^{n-1}\times$0.19 K km s$^{-1}$ corresponding to the $n$th level. The red contour levels of H41$\alpha$ emission are the same as those in Figure \ref{fig:H41a_center_width}. The dash-dotted lines presents the spatial resolution of $40''$.}\label{fig:SiO2-1_cv}
\end{figure}

The distributions of velocity-integrated intensity and central velocity of SiO 2-1 line are presented in Figure \ref{fig:SiO2-1_cv}. It is presented in the intensity distribution that the shocked gas indicated by the SiO line is distributed near the west and north boundaries of ionized region. The SiO line is most compact in the position near BGPS 3105. This is similar to the HCO$^+$ and H$^{13}$CO$^+$ 1-0 lines. In the top panel of Figure \ref{fig:HCO+_H13CO+}, the distributions of the SiO 2-1 and HCO$^+$ 1-0 lines are compared. The spreading area and even variation in intensity of the SiO 2-1 line are both approximately consistent with the compact emission of HCO$^+$ 1-0 line. This suggests that the shocked gas is probably stripped from the dense gas of molecular cloud. 

\begin{figure}
  \centering
  \includegraphics[scale=0.5]{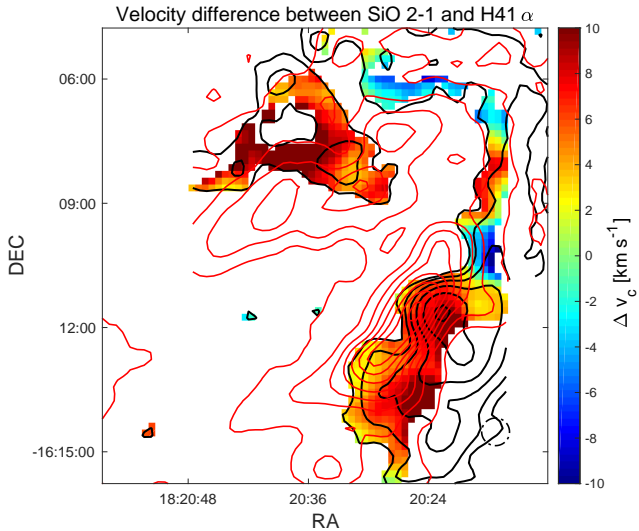}
  \caption{Distributions of difference between central velocities of SiO 2-1 and H41$\alpha$ lines. The contour levels are the same as those in Figure \ref{fig:SiO2-1_cv}.}\label{fig:H41a_SiO2-1_dv}
\end{figure}

The LOS velocity of shocked gas is presented by the central velocity of SiO 2-1 line. The variation of the central velocity of SiO 2-1 line is much weaker than that of H41$\alpha$ line. As that of HCO$^+$ 1-0 line,
the central velocity of SiO 2-1 line is about 18-20 km s$^{-1}$ in most parts of the observational field except in the region near BGPS 3114 where it increases to $\sim23$ km s$^{-1}$. In addition, the difference between central velocities of SiO 2-1 and H41$\alpha$ lines are shown in Figure \ref{fig:H41a_SiO2-1_dv}. The central velocity of SiO 2-1 line is more red-shifted than that of H41$\alpha$ in the two sides of the H II region. And this feature is not clear in the head. This implies the shocked gas is located behind the ionized gas in the sides of H II region, and the boundary of H II region is edge-on in the head.

\section{discussions} \label{sec:discussion}

The main contents of discussions are the relation between M17 H II region and adjacent clumps, and the kinetic characteristics of the ionized gas shown by the H41$\alpha$ line. The origin of shock in the boundary region of M17 H II region is studied in Section \ref{sec:origin_shock}. The further discussion of the formation of the massive clumps near M17 H II region is written in Section \ref{sec:origin_clump}. Derived from the hydrogen recombination line and continuum emission, some false SCCs are identified in Section \ref{sec:false_SCC}. In Section \ref{sec:stellar_motion}, evidences of one or more massive stars with supersonic velocity in M17 H II region are suggested according to the kinetic characteristics of ionized gas.

\subsection{Origin of the shock near M17 H II region} \label{sec:origin_shock}


Shocks near massive clumps could be originated from the protostellar outflows and collision of flowing gas. In general, the broad Gaussian component shown in the SiO spectrum is attributed to high-velocity shock from protostellar outflows, and the narrow Gaussian component is caused by less powerful outflows or collision between different clouds of gas \citep{lou16}. The FWHMs of the SiO line profiles toward protostellar sources in our observations are listed in Table \ref{table:proto_fwhm}. The the SiO 2-1 line profiles toward BGPS 3096, 3103 and 3105 can be fitted by the sum of a broad and a narrow Gaussian components. These broad SiO 2-1 components (FWHM$>9$ km s$^{-1}$) should be attributed to fast shock due to powerful protostellar outflows \citep{dua14,cse16}. Broad SiO 2-1 component is not clear in BGPS 3104 and 3124. This could be because the rms noise level is not low enough to show the high-velocity emission in the SiO 2-1 spectra toward these 2 sources.

However, the intensities in Table \ref{table:proto_fwhm} clearly indicate that narrow components are stronger than broad components even in these protostellar clumps. For SCCs, the FWHMs of the SiO 2-1 line profiles are also measured and found to be from 3 km s$^{-1}$ to 6 km s$^{-1}$. Shocked gas exists even away from these protostellar clumps. Hence, fast shock due to powerful protostellar outflows is not the dominant origin of the shocked gas near M17 H II region. Moreover, although less powerful outflows provided by intermediate- and low-mass protostars can leads to the narrow SiO emission, they can not produce the extended distribution of the SiO emission in the current observations unless a large population of undetected protostars are widespread in the boundary of M17 H II region. On the contrast, the collision between M17 H II region and molecular cloud is the most likely cause of the shocked gas since the compression of the nearby molecular cloud by M17 H II region has been found in previous works \citep{bro99,wil03}.

\begin{table} \tiny 
\centering
\caption{The central velocities, widths, and $\int T_{\textrm{mb}}dv$ of the SiO 2-1 line profiles toward protostellar clumps.}\label{table:proto_fwhm}
\begin{threeparttable}
\begin{tabular}{|c|ccc|}
\hline
source & $v_c$ [km s$^{-1}$] & FWHM [km s$^{-1}$] & $\int T_{\textrm{mb}}dv$ [K km s$^{-1}$] \\
\hline
\multirow{2}*{BGPS 3096} & $19.18\pm0.39$ & $7.23\pm1.78$ & $0.76\pm0.07$ \\
 & $11.73\pm6.10$ & $15.02\pm7.77$ & $0.60\pm0.09$ \\
\multirow{2}*{BGPS 3103} & $18.99\pm0.06$ & $4.82\pm0.18$ & $2.31\pm0.10$ \\
 & $13.61\pm0.99$ & $25.49\pm2.06$ & $2.12\pm0.17$ \\
BGPS 3104 & $20.83\pm0.17$ & $2.15\pm0.39$ & $0.22\pm0.04$ \\
\multirow{2}*{BGPS 3105} & $19.42\pm0.06$ & $8.25\pm0.23$ & $5.53\pm0.26$ \\
 & $17.48\pm1.44$ & $25.98\pm4.51$ & $1.51\pm0.27$ \\
BGPS 3124 & $21.08\pm1.75$ & $10.31\pm4.12$ & $0.39\pm0.11$ \\
\hline
\end{tabular}
\begin{tablenotes}
\item \textbf{Notes.} The SiO 2-1 line profiles in BGPS 3096, 3103, and 3105 can be divided into two components. The SiO 2-1 spectra are convolved with a $40''$ beam size to increase the signal-to-noise ratio.
\end{tablenotes}
\end{threeparttable}
\end{table}

Previous observations and simulations suggest that the collision between expanding ionized gas and ambient molecular gas can produce shock \citep{ten79,hos06,deh10,zhu15b}, and the ambient medium will fall into the shell of H II region after it is shocked.  As shown in Figures \ref{fig:HCO+_H13CO+} and \ref{fig:SiO2-1_cv}, the spatial distributions of the SiO and H41$\alpha$ lines show clear relation between the shocked and ionized gas. The brightness of the SiO emission is also correlated to the column density of molecular gas shown by the HCO$^+$ and H$^{13}$CO$^+$ 1-0 line intensities. So the shocked gas indicated by the SiO line should be mainly produced by the expanding ionized gas, and be stripped from the molecular cloud. 

\subsection{Relation between massive clumps and shock} \label{sec:origin_clump}

In \citet{hos06}, the shell of H II region is defined as the neutral layer between ionization and shock fronts. It is also called the cooled postshock (CPS) layer in \citet{elm77}, which suggested that the CPS layer of dense neutral material accumulates between the ionization and shock fronts and eventually becomes gravitationally unstable. This suggestion is also supported by the subsequent works in observations and simulations \citep{hos06,deh10}. For M17, \citet{elm77} pointed out that the ongoing star formation around molecular cloud core M17 SW is promoted by M17 H II region. In \citet{hof08}, A group of 647 accreting protostellar candidates identified with K-L colors were found in the periphery of massive cluster NGC 6618. This also supported the scenario of star formation in the nearby molecular cloud triggered by the expansion of M17 H II region.

Since the shocked gas near M17 H II region is mainly produced by the expanding ionized gas, the shocked gas should be located in the postshock layer surrounding the H II region. Although the FUV radiation from massive stars may destroy SiO molecules, this dissociating radiation can be much weakened in the CPS layer so that SiO molecules can survive in this layer if this layer is thick enough \citep{zhu15b}. So in our observations, the distribution of SiO emission presents the morphology and location of the CPS layer. In addition, the intensity distribution of the HCO$^+$ 1-0 line presents a dense layer of molecular gas along the boundary of the H II region. This dense layer is also overlapped by the SiO line emission and shows the brightest HCO$^+$ emission at every radial directions from the center of H II region to the outside. Hence, we suggest that this dense layer results from the projected image of the roughly edge-on part of CPS layer of M17 H II region.

Moreover, there are 21 massive clumps in the $\sim20\times20$ arcmin$^2$ area around M17 H II region \citep{gin13}. 10 clumps of them are located in the $\sim39$ arcmin$^2$ overlap region of the SiO 2-1 and HCO$^+$ 1-0 lines in the project image. There is a strong correlation between the spatial distributions of the CPS layer and the clumps near M17 H II region. If these clumps are actually embedded in the CPS layer, they should result from the accumulated dense material. It is not easy to affirm the formation origins of the clumps in molecular cloud core M17 SW. Some of them may be formed from star formation in the cloud core independent of the influence by M17 H II region \citep{hof08}. On the contrary, the 5 clumps including BGPS 3110, 3111, 3118, 3124, and 3128 are probable in the edge-on CPS layer. They seem to be formed under the influence of the H II region.

Derived from the 1.1 mm continuum \citep{gin13}, the masses, column densities, and volume densities of the clumps near M17 H II region are written in Table \ref{table:clumps}. The estimated gas densities in the massive clumps located in the overlap of SiO 2-1 and HCO$^+$ 1-0 emission mostly range from $10^4$ cm$^{-3}$ to $10^5$ cm$^{-3}$. In addition, \citet{whi94} also suggested that the CPS layer undergoes gravitational fragmentation before the column density reaches a critical value. Under the assumptions of kinetic temperature $T=25$ K, gas density $n_{\textrm{H}}=10^4$ cm$^{-3}$, and Lyman continuum photon flux $\sim 2.1\times10^{50}$ s$^{-1}$, the corresponding critical column density is $\sim3.0\times10^{22}$ cm$^{-2}$. The estimated column densities of BGPS 3096, 3102, 3103, and 3105 are higher than this value. Because these clumps are all in M17 SW, we can not conclude whether they are originated from fragments of the CPS layer or are pre-existing massive condensations due to star formation in the cloud core.

\subsection{Misclassified sources in the SCC sample} \label{sec:false_SCC}

The selected massive clumps in this work were identified by 1.1 mm continuum emission with a watershed decomposition algorithm \citep{gin13}. However, besides the thermal emission from heated dust in compact molecular clumps, the 1.1 mm continuum flux can also be the free-free emission produced by ionized gas. As presented in Figure \ref{fig:HCO+_H13CO+}, the starless candidates including BGPS 3100, 3107, 3108, 3114, 3115, and 3119 are not located in the compact regions of the HCO$^+$ and H$^{13}$CO$^+$ 1-0 lines but located in those of the H41$\alpha$ and 1.06 GHz continuum emission. It is possible that these SCCs are misclassified if the 1.1 mm continuum emission is mainly attributed to free-free emission in these sources. In order to check this speculation, the 1.1 mm continuum flux is compared with the 1.06 GHz continuum flux provided by \citet{beu16}.

In H II regions, the continuum flux contributed from heated dust is negligible at 1.06 GHz compared with that from free-free continuum emission \citep{gor02}. The brightness temperature $T_b$ of free-free continuum is approximately proportional to $\nu^{-2.1}$ when the continuum optical depth is thin \citep{alt60,gor02,pet12}. The estimated temperature of the ionized gas in M17 H II region is commonly higher than 7000 K \citep{hje71,dow80,sub96,tow03}. Except in the brightest position near BGPS 3105, the $T_{mb}$ of 1.06 GHz continuum is significantly lower than the estimated gas temperature in the whole M17 H II region. This suggests the optical depth of free-free continuum is thin at the frequency $\nu\geq1.06$ GHz. So the 1.1 mm free-free continuum intensity can be estimated from the 1.06 GHz continuum intensity. In Table \ref{table:misclassification}, the $40''$ flux density and corresponding $T_{mb}$ of 1.1 mm continuum from \citet{gin13}, and $T_{mb}$ estimated from 1.06 GHz continuum are listed. When estimating the 1.1 mm continuum $T_{mb}$, the spatial resolution of 1.06 GHz observation is convolved to be $40''$. Since the two series of values are roughly consistent in BGPS 3100, 3107, 3108, 3114, 3115, and 3119, these sources should not be the candidates of starless clumps. For BGPS 3100 and 3114, there are amounts of high-density gas indicated by HCO$^+$ and H$^{13}$CO$^+$ lines close to the sources. But since the properties and locations of BGPS 3100 and 3114 are estimated from the contaminated 1.1 mm continuum emission, we still regard them as false SCCs. \citet{gin13} claimed that the 1.1 mm continuum emission is unaffected by free-free emission. However, it seems easy to lead to misclassification of SCCs by using only millimeter and sub-millimeter continuum without dense-gas tracers. 

\begin{table} \tiny 
\centering
\caption{The misclassified sources in the SCC sample \citep{svo16}. The 1.1 mm continuum intensities ($T_{mb}$) from observations and from the estimation using 1.06 GHz continuum are compared. The 1.1 mm continuum intensities from observations are derived from the 40$''$ flux given by \citet{gin13}.}\label{table:misclassification}
\begin{tabular}{|c|ccc|}
\hline
Sources & 40$''$ flux & $T_{\textrm{mb}}$ from observations & $T_{\textrm{mb}}$ from estimation \\
\hline
BGPS 3100 & $1.25\pm0.13$ Jy & $1.86\times10^{-2}$ K & $2.08\times10^{-2}$ K \\
BGPS 3107 & $0.25\pm0.13$ Jy & $3.66\times10^{-3}$ K &  $4.26\times10^{-3}$ K \\
BGPS 3108 & $2.48\pm0.20$ Jy & $3.68\times10^{-2}$ K & $3.52\times10^{-2}$ K \\
BGPS 3114 & $2.19\pm0.16$ Jy & $3.25\times10^{-2}$ K & $3.26\times10^{-2}$ K \\
BGPS 3115 & $1.20\pm0.13$ Jy & $1.78\times10^{-2}$ K & $2.25\times10^{-2}$ K \\
BGPS 3119 & $1.63\pm0.14$ Jy & $2.41\times10^{-2}$ K & $2.68\times10^{-2}$ K \\
\hline
\end{tabular}
\end{table}

\subsection{The origins of the kinematic characteristics of ionized gas shown in the H41$\alpha$ line} \label{sec:stellar_motion}

M17 H II region is a giant cometary H II region. There are two series of models, champagne-flow models and bow-shock models, proposed to explain the formation of the cometary H II regions \citep{ten79,mac91,art06,zhu15b}. In champagne-flow models, the original density gradient in the natal molecular cloud causes the cometary morphology of H II region. In bow-shock models, the stellar motion with respect to the ambient medium is the dominant motivation. However, combined effects of these two models are found in the observation toward the southern H II region in DR 21 \citep{cyg03}. In M17 H II region, we suggest that the supersonic stellar motion and the density gradient are both present.

\subsubsection{The evidences for the density gradient and the stellar motion.}

The density gradient has been clearly confirmed in the Previous works \citep{wil03,pov09,sof22}. The molecular gas is compact in the west and north, and is sparse in the southeast. As presented in the intensity distributions of HCO$^+$ and H$^{13}$CO$^+$ 1-0 lines in Figure \ref{fig:HCO+_H13CO+}, our observations also show that the high-density gas is widespread in north and west, but lacking in the southeast. Moreover, the evidence for stellar winds from the central group of O stars is also strong \citep{tow03}. However, a single champagne-flow model with a stellar wind has several difficulties to explain some features shown in the H41$\alpha$ line.


1. As listed in Table \ref{table:global_property}, the central velocity of the H41$\alpha$ line toward the whole H II region is $19.0$ km s$^{-1}$ very close to the $\sim20.0$ km s$^{-1}$ center velocity of M17 molecular cloud measured in CO 1-0 line \citep{sof22}. Such a little difference in central velocity can not be produced by champagne-flow models \citep{zhu15b} except when the line of sight is roughly perpendicular to the symmetry axis of the cometary H II region. But an inclination angle of $90^o$ can not explain the clearly asymmetrical feature in the P-V diagram in Figure \ref{fig:H41a_pvdiagram}.


2. A zone composed of significantly red-shifted ionized gas is distributed along the symmetry axis and from the center to the southeast of M17 H II region. For an axial symmetrical cometary H II region, the large amount of red-shifted ionized gas in this zone with $> 10$ km s$^{-1}$ velocity relative to the molecular gas can not be explained by the champagne-flow model \citep{zhu15b}. Especially for the considerable H41$\alpha$ component with velocity $> 30$ km s$^{-1}$ shown in Figure \ref{fig:H41a_pvdiagram}, this feature seems to exist only in bow-shock models. The fast expanding stellar wind bubble and photoionized region in the beginning of a young compact H II region may lead to this velocity by pushing ionized gas away from the center massive star, but M17 H II region is a giant H II region with an old age.

3. As plotted in Figure \ref{fig:H41a_spectrum}, the H41$\alpha$ spectra can all be divided into two Gaussian components with different velocities. The central velocity of the red-shifted component in these spectra decreases from the head to the tail. This also suggests the existence of high stellar velocity relative to the natal molecular cloud since the velocity of photoionized gas also gradually decreases from the head to the tail in bow-shock models \citep{zhu15b}.

Then we suggest that at least one massive star in M17 H II region should have a supersonic stellar velocity with respect to the molecular cloud.

\subsubsection{Simulation of a combined model including both the density gradient and the stellar motion}

Using the 2-D dynamical model of cometary H II region with cylindrical coordinates ($r$, $z$) \citep{zhu15b}, the existences of gas density gradient and supersonic stellar motion are checked with the result of a simulation. Here $z$ is the axial coordinate pointing from the outside to the center of the molecular cloud. $r$ is the radial coordinate. The simulation is computed on a $250\times500$ grid. The cell size of the grid is 0.02 pc. In the beginning of the evolution, the gas density $n_{\textrm{H}}(z)$ and temperature $T(z)$ are $10^4$ cm$^{-3}$ and 20 K for $z>0$ pc, respectively. This gas density is suggested as the initial density of M17 molecular cloud \citep{sof22}. The temperature is a typical value for molecular cloud \citep{dra11}. For $z<-1$ pc, the initial values of $n_{\textrm{H}}(z)$ and $T(z)$ are $20$ cm$^{-3}$ and 100 K, respectively. They are reasonable for cold neutral medium \citep{dra11}. It is necessary to note that the exact density and temperature of the diffuse gas out of molecular cloud are not important in this simulation. The pressure of the diffuse gas is much lower than that of ionized gas so that its effect on the evolution of H II region can be neglected. In the area with $-1$ pc $\leq z \leq$ 0 pc, the gas density follows an exponential law as $n_\textrm{H}(z)=n_0 exp(z/H)$. The number density $n_0$ is $10^4$ cm$^{-3}$, and the scale height $H$ is 0.4 pc. Exponential density distribution with a scale height is a good approximation to the density gradient in a molecular cloud \citep{art06}. In this simulation, the gradual decrease of density from $z=0$ pc to $-1$ pc is used to imitate the decreasing density in the boundary region of molecular cloud. The temperature in the boundary region also follows an exponential law to keep the pressure equilibrium although the equilibrium will be broken quickly after the simulation begins. The heating gained per photoionization is assumed to be 2 eV in the calculation. This makes the electron temperature in photoionized region is $T_e\sim 7000$ K roughly like $T_e$ in M17 H II region. The gas velocity in the whole grid is set to be 0 km s$^{-1}$ in the beginning. And the H II region is assumed to be observed from the tail to the head with an inclination angle of $30^o$ between the symmetrical axis and the line of sight as in \citet{zhu15b}.

There are dozens of ionizing stars located in M17 H II region, but they are simplified into three stars in the simulation. Two stars with $2.0\times10^{50}$ s$^{-1}$ and $0.5\times10^{50}$ s$^{-1}$ photon luminosities of ionizing radiation ($h\nu\geq 13.6$ eV) are located at ($r$, $z$)$=$(0, 0) and (0, 1) in the start of simulation. They are mainly responsible for the ionization in the H II region, but have no stellar winds and no stellar velocities. The Lyman continuum photon flux contributed from ionizing stars in M17 H II region can be estimated from the continuum flux \citep{woo89}. We calculate the continuum flux at $\nu=92.034$ GHz from the observed H41$\alpha$ line flux with the departure coefficients estimated under the assumptions of $T_e=7000$ K and $n_e=10^3$ cm$^{-3}$ \citep{zhu19,zhu22}. And then the Lyman continuum photon flux is estimated to be $2.1\times10^{50}$ s$^{-1}$. In this simulation, we increase the total ionizing photon luminosity by about $30\%$ because a part of ionizing photons might escape through the tail of H II region, and the tail of H II region is not totally covered in our observation.

A cometary H II region is immediately produced due to the gas gradient after the simulation starts. When the evolutionary age is 20,000 yrs, an O4 V star with a stellar wind is assumed to move into the grid. The stellar velocity is $v_*=25$ km s$^{-1}$ with a direction along the symmetrical axis from the outside to the center of molecular cloud. The mass-loss rate and the terminal velocity of the stellar wind are $\dot{M}=5\times10^{-6}$ M$\odot$ yr$^{-1}$ and $v_w=2979$ km s$^{-1}$, respectively \citep{han97,dal13,kob19}. The photon luminosity of ionizing radiation of this moving star is $2.75\times10^{49}$ s$^{-1}$ \citep{han97}. The formation of stellar wind bubble is attributed to this star.

\begin{figure}
  \centering
  \includegraphics[scale=0.45]{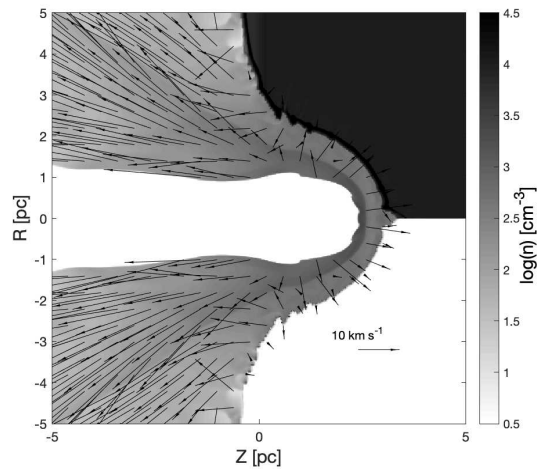}
  \caption{Density distributions of all materials (top half), ionized gas (bottom half) at the age of 450,000 years. The arrows represent velocity field in photoionized region.}\label{fig:model_map}
\end{figure}

The simulation is stopped at 450,000 yrs when the size of the H II region reaches about that of M17 H II region. The density distribution is presented in Figure \ref{fig:model_map}. The low-density zone around symmetrical axis is the stellar wind bubble. The hot and low-density bubble is surrounded by the photoionized gas whose velocity field is also shown. It is clearly presented that the velocity of ionized gas is red-shifted ($v>0$) in the head and blue-shifted ($v<0$) in the sides and tail of the cometary H II region. In addition, the thickness of the shocked neutral gas in the stimulation is about 0.15 pc ($15.5''$). This is roughly consistent with the thickness of the shocked gas shown by the SiO line. Then the hydrogen RRLs emitted from the H II region is simulated. The details of the simulation of RRLs are given in Appendix \ref{sec:method_rrl}.

\begin{figure}
  \centering
  \includegraphics[scale=0.43]{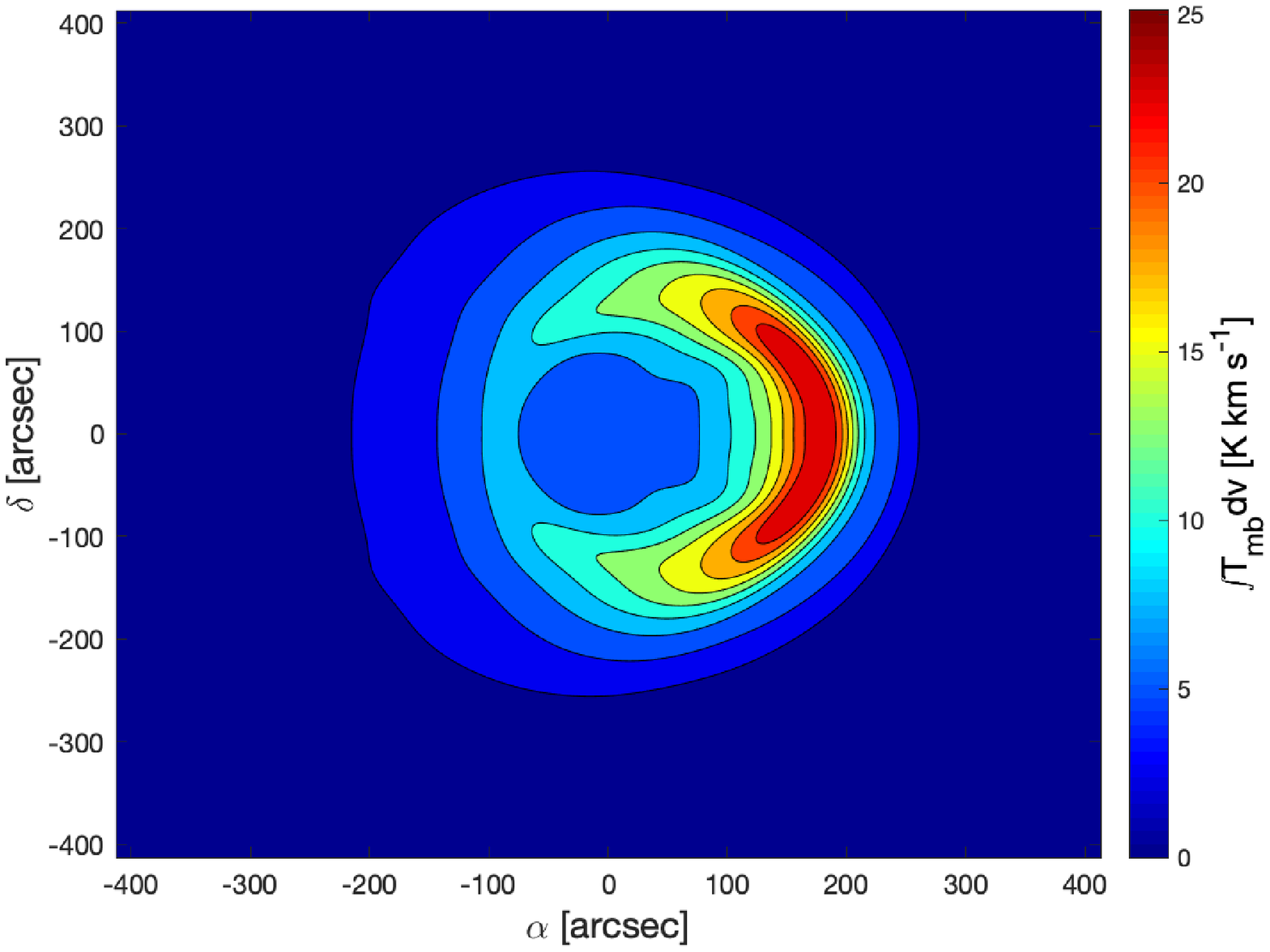}
  \includegraphics[scale=0.43]{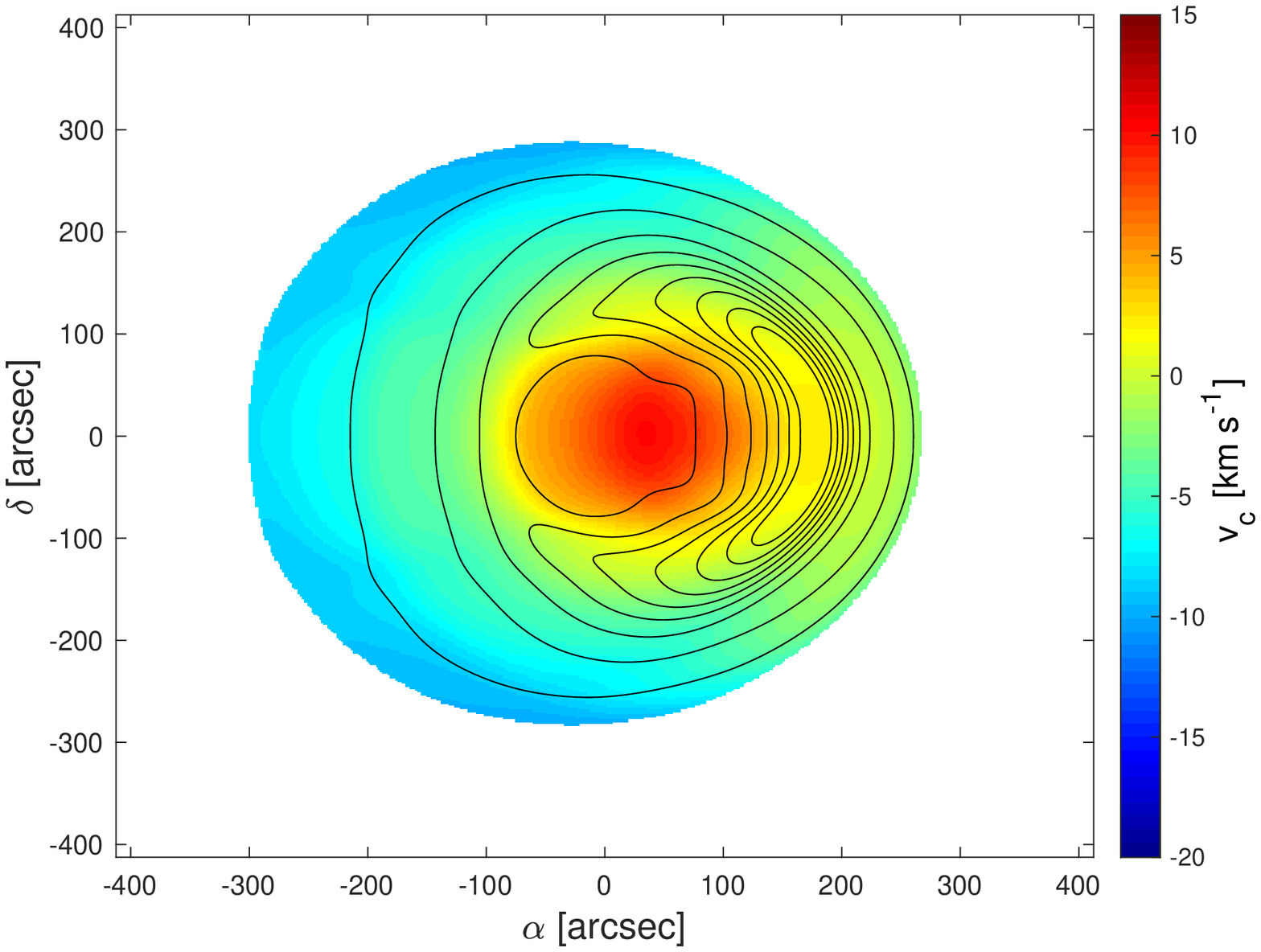}
  \includegraphics[scale=0.43]{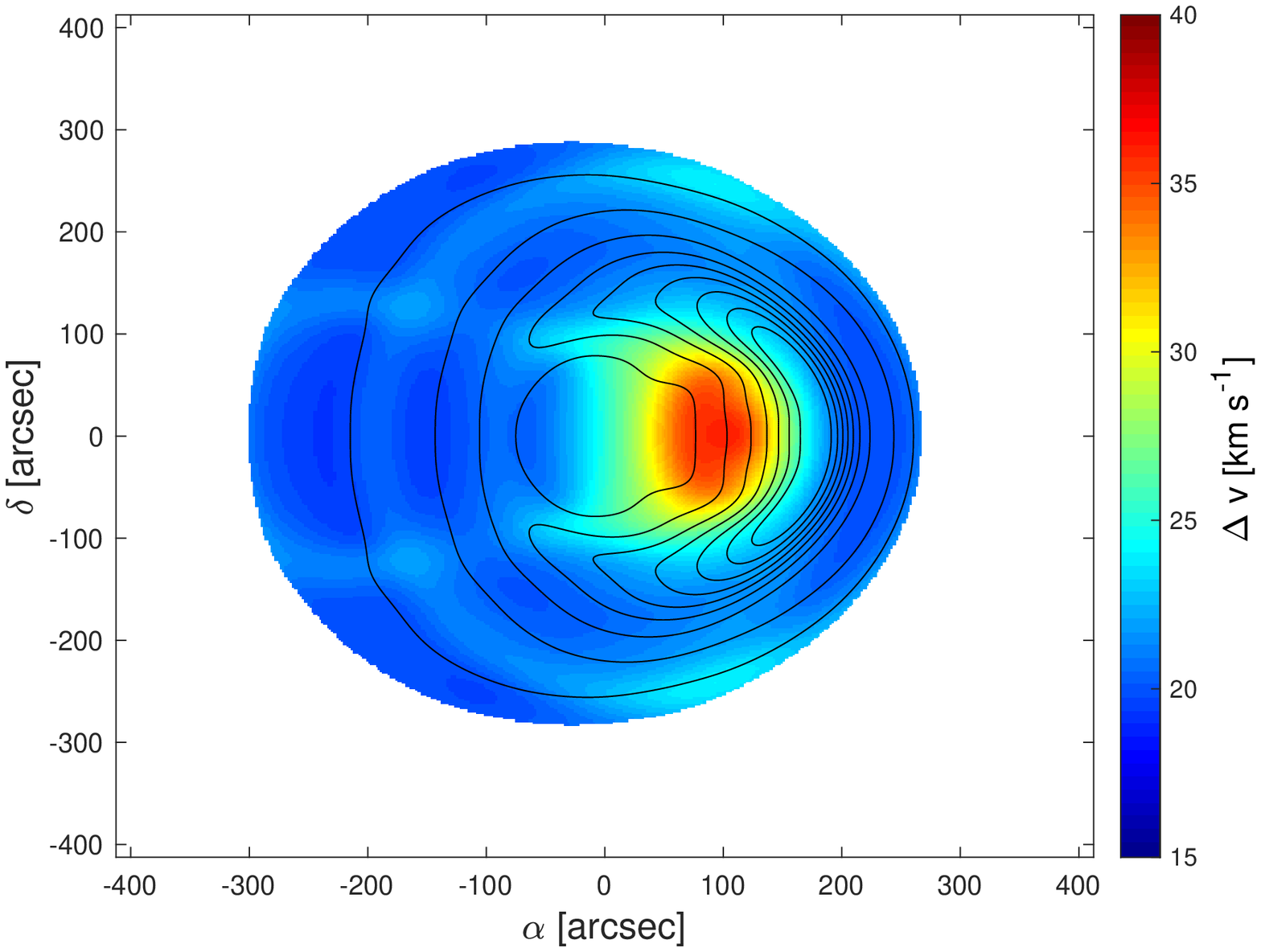}
  \caption{Distributions of velocity-integrated intensity (top panel), central velocity (middle panel), $\Delta v$ (bottom panel) of the H41$\alpha$ line in the simulation. The inclination angle is $30^o$. The simulated beam size is $30''$. The distance of the simulated H II region is assumed to be 2 kpc as the distance of M17 H II region. The contour levels starts at 0.1 maximum value of velocity-integrated intensity in steps of 0.1 maximum value. The central velocity and $\Delta v$ are only shown in the position where the intensity is higher than 0.05 maximum value.}\label{fig:model_H41a}
\end{figure}

The projected 2-D distributions of velocity-integrated intensity, central velocity, width of the H41$\alpha$ line are shown in Figure \ref{fig:model_H41a}. The simulated H II region is assumed to be 2 kpc distant, and the simulated beam size is $30''$. The simulated velocity-integrated intensity of the H41$\alpha$ line is presented in the top panel. The morphology of the compact region is similar to that displayed in Figure \ref{fig:H41a_ctn}. Although it is different from the observation that the H41$\alpha$ line emission is brightest in the apex, the arch-shaped compact emission and the faint zone in the center are reproduced in the simulation. It is necessary to notice that the arch-shaped compact emission mainly results from the projection of the line emission from the photoionized region. For the southern and northern bars of M17 H II region, we suggest a large part of the line and continuum emission is attributed to the projection effect of the sides of H II region although there are some individual ultra-compact H II regions which contribute a part of emission.

The central velocity distribution of H41$\alpha$ line is shown in the middle panel. The distribution of central velocity is very similar to that of the observation shown in the top panel of Figure \ref{fig:H41a_center_width}. The central velocity of ionized gas is red-shifted in the faint zone which is in the center and half-surrounded by the arch-shaped compact region. And the central velocity gradually decreases to the sides and tail.

The distribution of line width shown in the bottom panel of Figure \ref{fig:model_H41a} seems to be not very like that in the observation. The layer with $\sim35$ km s$^{-1}$ broad H41$\alpha$ line width ($\Delta v$) around the red-shifted gas is not formed in the simulation although there is a mechanism to increase the line width slightly to $\sim23$ km s$^{-1}$ in the layer around the stellar wind bubble. In the observation, the line width in the layer is significantly broader than the line width $\Delta v<28$ km s$^{-1}$ in the surrounded center zone. This phenomenon is not found in the simulation. Although the H41$\alpha$ line width could be underestimated in the simulation because the microturbulence is not considered, this can not explain the differences between the observation and simulation mentioned above. However, in the simulation, the broad line profile in the position ahead of the red-shifted ionized gas is realized corresponding to the broad H41$\alpha$ emission in Region A. 

\begin{figure}
  \centering
  \includegraphics[scale=0.4]{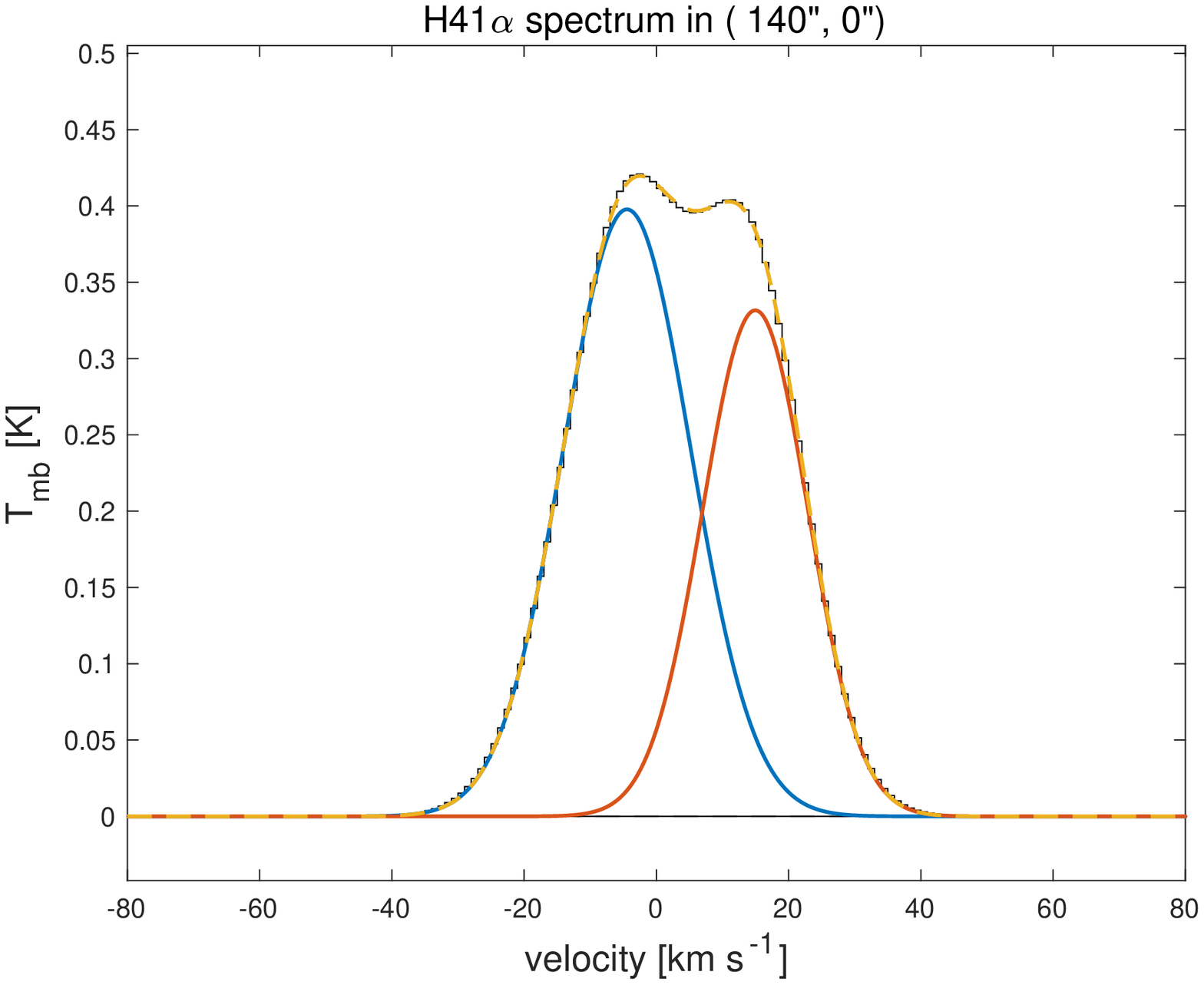}
  \includegraphics[scale=0.4]{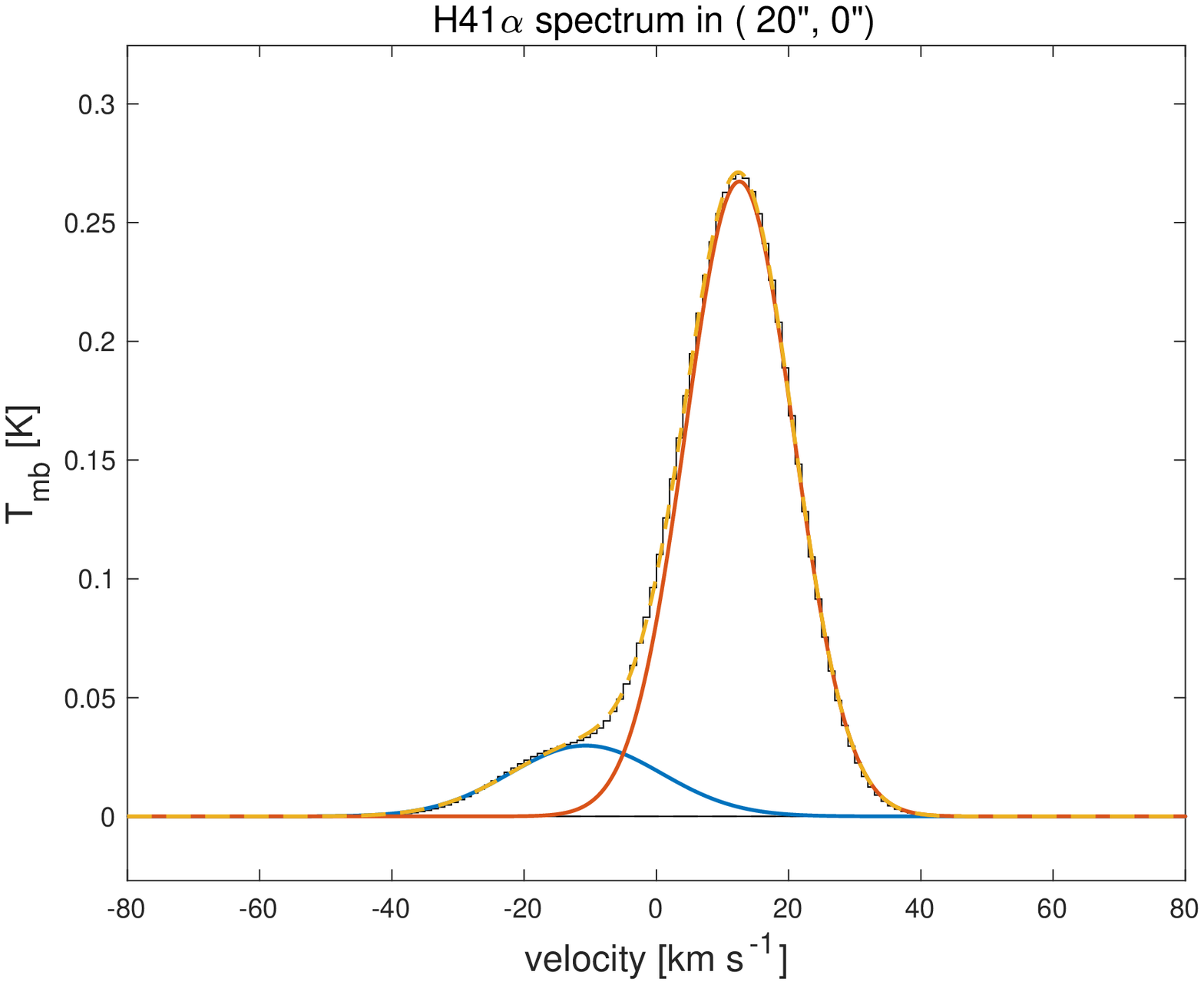}
  \caption{The H41$\alpha$ spectra toward the head and the center of the simulated H II region. The profiles of these spectra are similar to those in the top and bottom panels of Figure \ref{fig:H41a_spectrum}. The coordinates ($\alpha$, $\delta$) are corresponding to those in Figure \ref{fig:model_H41a}.}\label{fig:model_spec}
\end{figure}

The simulated H41$\alpha$ spectra toward the head and the center of the H II region are plotted in Figure \ref{fig:model_spec}. The double-Gaussian profiles are similar to those in the top and middle panels of Figure \ref{fig:H41a_spectrum} although the line intensity is higher in the simulation than in the observation. The blue-shifted velocity component is attributed to the ionized gas in the sides of H II region while the red-shifted one is contributed from the ionized gas pushed by the stellar wind of the moving star.



The real environment in M17 H II region should be very complicated. A 2-D cylindrical model calculated with assumptions on massive stars and the molecular cloud is too simple to reproduce every features in the observations. The line emission is brightest in the apex in the simulation but is relatively weaker than southern and northern bars in the observation. It may be caused by a non-cylindrical distribution in the real environment. If the ionized gas is more diffuse in the side facing us, this feature of the observed H41$\alpha$ line could be realized. This needs to be checked in a 3-D model.


However, since many features of the H41$\alpha$ line emission in our observations are realized in the simulation, we think that the combined model including both of champagne flow and bow shock can explain the morphology and kinematics of M17 H II region. Although they are not presented in this work, the cases with different stellar velocities have also been simulated. It is found that a stellar velocity $v_*\geq25$ km s$^{-1}$ is necessary to produce the high-velocity red-shifted gas shown in our observation. Then we suggest that there is one or more O stars moving to the molecular cloud with a supersonic stellar velocity higher than 25 km s$^{-1}$ in M17 H II region.  


\section{Summary} \label{sec:conclusion}

In this paper, we present the observations of the SiO 2-1, HCO$^+$ 1-0, H$^{13}$CO$^+$ 1-0, HC$_3$N 10-9, and H41$\alpha$ lines toward M17 H II region and the adjacent massive clumps. The overall properties of these radio lines in M17 H II region are shown. The intensity, central velocity and width distributions of these lines are displayed. The properties of the relevant molecular and ionized gas are studied. The observations toward M17 H II region are compared with the simulation of dynamical model of cometary H II region in order to study the morphology and kinematics of ionized gas in M17 H II region. The conclusions are summarized as follows.

1. It is found that the shocked gas spreads near the boundary of M17 H II region. This suggests that this widespread shock should be originated from the expansion of ionized gas.

2. There could be some massive clumps located in the CPS layer between the shock and ionization fronts since they are covered by the SiO 2-1 line emission. The number density of massive clumps in the CPS layer indicated by the SiO 2-1 and HCO$^+$ 1-0 emission in the projected map is  $\sim 0.26$ arcmin$^{-2}$ while that of massive clumps in the vicinity of M17 H II region is $\sim 0.05$ arcmin$^{-2}$. The correlation between massive clumps and the CPS layer of M17 H II region is clear. The accumulation of swept-up neutral materials in the CPS layer of H II region should be a main origin of these clumps.

3. From the 1.06 GHz continuum intensity given by \citet{beu16}, the 1.1 mm continuum intensities due to free-free emission from the positions of the SCCs overlapped with ionized gas in M17 H II region are derived. A number of false SCCs including BGPS 3100, 3107, 3108, 3114, 3115 and 3119 are identified.

4. The formation of the cometary H II region and its kinematics should be caused by the both effects of a moving massive star and an initial density gradient between the inside and outside of molecular cloud. The density gradient is clearly shown in previous works \citep{sof22}, and is also supported by the spatial distributions of HCO$^+$, H$^{13}$CO$^+$ 1-0, and HC$_3$N 10-9 lines in the current observations. From the comparison between our observations and simulations, a massive star with a strong stellar wind and a supersonic velocity $v_*\geq25$ km s$^{-1}$ is suggested in M17 H II region.

\section*{Acknowledgements}

This work is based on the observations performed using IRAM 30m telescope. It is supported by the National Science Foundation of China No. 12003055, and Key Research Project of Zhejiang Lab (No. 2021PE0AC03). Y. T. Y. is a member of the International Max Planck Research School (IMPRS) for Astronomy and Astrophysics at Universities of Bonn and Cologne. Y. T. Y. would like to thank China Scholarship Council (CSC) and the Max-Planck-Institiut f\"{u}r Radioastronomie (MPIfR) for the financial support.

\section*{DATA AVAILABILITY STATEMENT}

The data underlying this paper will be shared on reasonable request to the corresponding author.

\begin{appendix}

\section{Method of simulating hydrogen RRLs} \label{sec:method_rrl}

The simulations of the evolution of H II region are calculated by the dynamical model of the H II region created by \citet{zhu15a} and \citet{zhu15b}. After the distributions of the physical and chemical quantities are calculated from the model of H II region, the model of calculating hydrogen recombination lines \citep{zhu19,zhu22} is used to simulate the intensity distributions of hydrogen recombination lines and continuum emission. Derived from the properties of the calculated hydrogen recombination lines with a beam size of $\sim30''$,  the observable quantities in real observations are simulated. Then compared with the results in simulations, the properties of ionized gas and even neutral gas traced by the H41$\alpha$ and molecular lines in the observations can be analyzed.

\section{Masses, column densities, and volume densities of clumps} \label{sec:clump_properties}

The masses, column densities, and volume densities of the clumps near M17 H II region are listed in Table \ref{table:clumps}. The masses of clumps are estimated from the total flux densities of 1.1 mm continuum emission, and the H$_2$ column densities are estimated from the 1.1 mm continuum 40$''$ flux densities \citep{gin13}. The methods of estimating masses and column densities are the same as those in \citet{zhu20}. The volume densities are estimated from the masses and the radii of the clumps under the assumption of spherical symmetry. The radii of clumps are also given by \citet{gin13}. The kinetic temperatures of gas in clumps used in estimation are provided by \citet{svo16}. For the clumps without kinetic temperatures, a temperature of 26 K is assumed. This value is the average temperature of the clumps with measured temperatures near M17 H II region. The distance of all the clumps is assumed to be 2 kpc. The little difference of adopted distance leads to slightly different clump masses between this work and \citet{zhu20} for BGPS 3110, 3118, and 3128. The different assumed kinetic temperature of BGPS 3114 also produce the lower mass and column density in this work. In Section \ref{sec:false_SCC}, it is indicated that the continuum emission in BGPS 3100, 3107, 3108, 3114, 3115, and 3119 should be mostly contributed by the free-free radiation from ionized gas. However, the continuum emission is still assumed to be all attributed to thermal dust continuum in the calculation of masses, column densities, volume densities in Table \ref{table:clumps},

\begin{table} \tiny 
\centering
\caption{The masses, column densities, and volume densities of clumps derived from the 1.1 mm continuum intensity \citep{gin13}. }\label{table:clumps}
\begin{tabular}{|c|ccc|}
\hline
Sources & Mass [M$_\odot$] & column density [$10^{22}$ cm$^{-2}$] & density [$10^{4}$ cm$^{-3}$] \\
\hline
BGPS 3096 & $1700\pm504$ & $3.55\pm1.06$ & $1.34\pm0.40$ \\
BGPS 3100 & $523\pm153$ & $1.30\pm0.40$ & $2.48\pm0.74$ \\
BGPS 3102 & $1797\pm531$ & $5.76\pm1.72$ & $12.7\pm3.8$ \\
BGPS 3103 & $2610\pm767$ & $6.44\pm1.90$ & $3.37\pm1.00$ \\
BGPS 3104 & $2246\pm656$ & $2.46\pm0.74$ & $2.46\pm0.73$ \\
BGPS 3105 & $8530\pm2497$ & $18.8\pm5.5$ & $7.30\pm2.14$ \\
BGPS 3107 & $55\pm21$ & $0.26\pm0.16$ & $0.27\pm0.10$ \\
BGPS 3108 & $469\pm138$ & $2.59\pm0.77$ & $8.02\pm2.34$ \\
BGPS 3110 & $380\pm112$ & $1.77\pm0.54$ & $4.42\pm1.30$ \\
BGPS 3111 & $630\pm186$ & $1.05\pm0.34$ & $1.27\pm0.38$ \\
BGPS 3114 & $2442\pm717$ & $2.28\pm0.68$ & $1.32\pm0.39$ \\
BGPS 3115 & $404\pm119$ & $1.25\pm0.39$ & $1.00\pm0.30$ \\
BGPS 3118 & $377\pm113$ & $0.97\pm0.31$ & $1.16\pm3.47$ \\
BGPS 3119 & $1046\pm311$ & $1.70\pm0.51$ & $1.35\pm0.40$ \\
BGPS 3124 & $257\pm77$ & $0.75\pm0.24$ & $1.56\pm4.66$ \\
BGPS 3128 & $165\pm51$ & $0.91\pm0.29$ & $1.42\pm0.43$ \\
\hline
\end{tabular}
\end{table}

\end{appendix}






\bsp	
\label{lastpage}
\end{document}